\documentclass[aps,prb,a4,superscriptaddress,preprintnumbers,
showpacs,twoside,twocolumn,floatfix]{revtex4}
\usepackage{amsfonts}

\usepackage{graphicx}     
\usepackage{bm}
\usepackage{bm}
\usepackage{graphicx}
\usepackage{amsfonts}
\usepackage{amsmath}
\usepackage{amssymb}
\usepackage{amsthm}
\usepackage{bbm}
\usepackage{setspace}
\usepackage{helvet}
\usepackage{times}

\begin{document}
\title{Magnetic Bose glass phases of
coupled antiferromagnetic dimers with site dilution}
\author{Rong Yu}
\affiliation{Department of Physics \& Astronomy, Rice University, Houston, TX 77005, USA}
\author{Omid Nohadani}
\affiliation{School of Industrial Engineering, Purdue University, West Lafayette, IN 47907, USA}

\author{Stephan Haas}
\affiliation{Department of Physics \& Astronomy, University of
Southern California, Los Angeles, CA 90089-0484, USA}
\author{Tommaso Roscilde}
\affiliation{Laboratoire de Physique, \'Ecole Normale Sup\'erieure de Lyon,
46 All\'ee d'Italie, 69003 Lyon, France}


\date{\today}

\begin{abstract}

We numerically investigate the phase diagram of two-dimensional site-diluted
coupled dimer systems in an external magnetic field. We show that
this phase diagram is characterized by the presence of an extended
\emph{Bose glass}, not accessible to mean-field approximation, and
stemming from the localization of two distinct species of bosonic
quasiparticles appearing in the ground state. On the one hand,
non-magnetic impurities doped into the dimer-singlet phase of a
weakly coupled dimer system are known to free up local magnetic
moments. The deviations of these local moments from full polarization
along the field can be mapped onto a gas of bosonic quasiparticles,
which undergo condensation in zero and very weak magnetic fields,
corresponding to transverse long-range antiferromagnetic order.
An increasing magnetic field lowers the density of such quasiparticles
to a critical value at which a quantum phase transition occurs,
corresponding to the quasiparticle localization on clusters of local magnets
(dimers, trimers, etc.) and to the onset of a Bose glass. Strong
finite-size quantum fluctuations hinder further depletion of
quasiparticles from such clusters, and thus lead to the appearance
of \emph{pseudo-plateaus} in the magnetization curve of the system.
On the other hand, site dilution hinders the field-induced
Bose-Einstein condensation of triplet quasiparticles on the intact
dimers, and it introduces instead a Bose glass of triplets. A
thorough numerical investigation of the phase diagram for a planar
system of coupled dimers shows that the two above-mentioned
Bose glass phases are continuously connected, and they overlap in a
finite region of parameter space, thus featuring a two-species Bose
glass. The quantum phase transition from Bose glass to magnetically ordered
phases in two dimensions is marked by novel universal exponents
($z\approx2$, $\beta\approx 0.9$, $\nu\approx 1$). Hence we conclude that doped 
quantum antiferromagnets in a field represent an ideal setting for the
study of fundamental dirty-boson physics.
\end{abstract}
\maketitle

\section{Introduction}\label{s.Introduction}
Quantum phase transitions in spin-gapped antiferromagnets (AF) have
been extensively studied both theoretically and experimentally.
Several mechanisms exist which can drive these systems from
a quantum disordered ground state with a finite spin gap to
a gapless antiferromagnetic long-range ordered (AFLRO) state. Possible
ways to close the spin gap are the modulation
of the superexchange couplings by, \emph{e.g.},
mechanical deformation of the crystal under pressure
\cite{Gotoetal06}, the application of a
magnetic field~\cite{Rice02, Zheludev05, Hondaetal,
Chaboussantetal97, Watsonetal01, Jaimeetal04, Sebastianetal05,
Nikunietal00, Rueggetal03, Zapfetal06, Klanjseketal08} or the dilution of the system with
non-magnetic impurities~\cite{Uchiyamaetal99, Azumaetal97,
Oosawaetal02}.

For definiteness we will discuss the case of spin-gapped systems
with a singlet ground state, associated with
the magnetic Hamiltonian of weakly coupled dimers. In these
systems a finite singlet-triplet spin gap opens for sufficiently weak
inter-dimer couplings. The application of a magnetic field
to the system can close the gap at a critical value at
which a quantum phase transition occurs, which is
well described as a Bose-Einstein condensation (BEC) of $S=1$
triplets~\cite{Nikunietal00, Giamarchietal08, Affleck91,
GiamarchiT99, Matsumotoetal04, Kawashima04, MisguichO04,
Wesseletal}. The field has the combined effects of partially
polarizing the dimers and inducing transverse AFLRO,
which corresponds to a superfluid (SF) state of the field-induced
triplets.
Alternatively, doping weakly-coupled dimer systems
with static non-magnetic impurities also has dramatic effects.
Each dopant releases a spin-$1/2$ degree of freedom out
of the singlet, and this $S=1/2$ moment
gets exponentially localized around the
site of the spin left unpaired~\cite{Sorensen98}. Hereafter
this localized $S=1/2$ degree of freedom will be denoted
as \emph{local moment} (LM).
The overlap of the localized wave functions of two nearby LMs
produces effective couplings that decay
exponentially with the distance between the two
impurities~\cite{SigristF96}, and which alternate
in sign depending on whether the LMs belong
to the same sublattice or not, and consequently they
are not frustrated. The resulting random network of LMs
therefore develops AFLRO at $T=0$ for any finite doping
concentration \cite{Laflorencieetal04}, and its
low-energy excitations 
give rise to a phenomenon of order-by-disorder
(OBD)~\cite{ShenderK91,Wesseletal01}.

It is intriguing to investigate the nature of the ground state in a
spin-gapped system when \emph{both} factors perturb the gapped
state, \emph{i.e.}, a magnetic field and non-magnetic impurities, are
present. In this paper we focus on the interplay between site
dilution and an applied magnetic field in a planar coupled dimer
system. We provide a detailed picture of the rich phase diagram
of the system, with particular emphasis on the emergence of an
extended novel quantum-disordered phase with the nature of a Bose glass
(BG), as introduced for three-dimensional systems in
Refs.~\onlinecite{Nohadanietal05, Yuetal10} and for two-dimensional systems 
in Refs.~\onlinecite{RoscildeH05,RoscildeH06,Roscilde06}. 
In particular we concentrate on the appearance of
a disordered-local-moment (DLM) phase appearing in the
diluted system upon application of a weak magnetic field
\cite{Roscilde06, Yuetal08}. This
phase is characterized by a highly inhomogeneous response of the
network of LMs, with a portion of the network (made of nearly
isolated LMs) being completely polarized by the field, and another
substantial remaining portion of interacting LM
clusters which takes a partially polarized state, namely it hosts
localized magnetic ``holes" in the background of LMs aligned with the
field. Due to the discrete structure of the clusters supporting the
holes (namely dimers, trimers etc. of LMs), the magnetization
process in the DLM phase proceeds in smooth steps: its
characteristic feature is represented by pseudo-plateaus (PPs)
associated with the magnetic response of small LM clusters subject
to a broad distribution of local fields exerted on them by the
remaining polarized LMs.

We here list the main results of this paper. \\
\indent
1) We quantitatively link the
features of the magnetization curve to the statistics
of the LM clusters and the effective local fields.
We compare the results for a site-diluted coupled-dimer
system with those for an effective model of a random
network of LMs, both in the quantum case of $S=1/2$
and in the classical ($S=\infty$) limit.
The random network of quantum LMs characterizes
the DLM phase, while the comparison
with the classical model highlights the strong
quantum nature.\\
\indent
2) We introduce a spin-to-boson mapping
which leads to a straightforward 
interpretation of the magnetic holes hosted
by the LM clusters as
localized bosonic quasi-particle states, whose
population can be changed by infinitesimal
changes of the applied field (namely the quasi-particle
chemical potential). Hence the resulting DLM phase is
a compressible disordered phase, analogous to a
Bose glass. \\
\indent
3) We map out the entire phase diagram
of the system for various inter-dimer couplings,
showing strong deviations from the mean-field
phase diagram of Ref.~\onlinecite{Mikeskaetal04}.
The phase diagram shows that the DLM phase is indeed
continuously connected to a higher field BG phase.
A detailed quantum-critical scaling analysis reveals
consistent critical exponents for the
OBD-to-DLM transition and for the BG-to-SF transition.
Hence we provide numerical evidence of a novel universality
class associated with the BG transition in two
dimensions.

The paper is organized as follows. In Sec.~\ref{s.Models} we
describe the models investigated and the numerical tools used in this work,
In Sec.~\ref{ss.PD_weak}, we briefly
review the numerical results exhibiting the sequence of quantum phases and
quantum phase transitions induced by the field in the 2D site-diluted
coupled-dimer system. In Sec.~\ref{s.QDPhase} we discuss
the structure of the magnetic response
and energetics in the DLM phase, and
we quantitatively explain the main features of the DLM phase. 
In Sec.~\ref{s.PD_entire}
we discuss the phase diagram of the 2D site-diluted dimer system as a function of
inter-dimer coupling strengths in the presence of a magnetic field.

\begin{figure}[h]
\begin{center}
\includegraphics[
     width=60mm,angle=0]{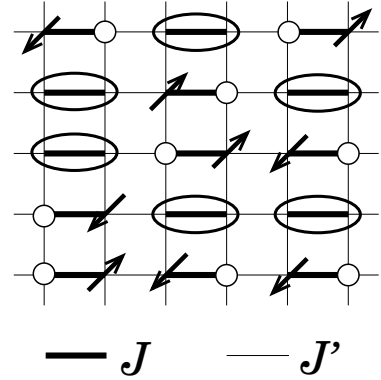}
\caption{2D coupled-dimer model with site dilution.
The ellipses indicate the formation of local singlets
on the intact dimers, open circles indicate vacancies,
and arrows represent local moments.}
\label{f.model}
\end{center}
\end{figure}

\section{Models and Numerical Method}\label{s.Models}

We consider a two-dimensional $S=1/2$ Heisenberg
antiferromagnetic model of weakly coupled dimers on a square
lattice.~\cite{Yasudaetal01B, Matsumotoetal01} The Hamiltonian reads
\begin{eqnarray}
{\cal H} &=&J\sum_{\langle ij \rangle \in {\rm dimers}}\epsilon_i \epsilon_j
{\bm S}_{i}\cdot
{\bm S}_{j}
+J'\hspace{-.8cm}\sum_{\langle lm \rangle \in {\rm inter-dimers}}
\hspace{-.8cm}\epsilon_{l}\epsilon_{m}
{\bm S}_{l}\cdot
{\bm S}_{m} \nonumber\\
&-&h\sum_{i}\epsilon_i \emph{S}^z_{i}, \label{e.ham_dilute}
\end{eqnarray}
where $\epsilon$ takes the values 0 or 1 with probabilities $p$ and
$(1-p)$, modeling the doping of the system with non-magnetic
impurities.  The pairs $\langle ij \rangle$ run on the strong
dimer couplings $J$, and the pairs $\langle lm \rangle$
on the weaker inter-dimer couplings $J'$ - see Fig.~\ref{f.model}.
In Sec.~\ref{ss.PD_weak} and ~\ref{s.QDPhase} we 
first focus on the weak-coupling limit, i.e., $J'\ll J$. Later, in
Sec.~\ref{s.PD_entire} we investigate the entire phase diagram for
$J'\lesssim J$.

In the clean limit ($p=0$) and at zero magnetic field there is a
critical value $(J'/J)_c \simeq 0.523$ separating the magnetically
ordered phase for $J'/J>(J'/J)_c$ from a dimer-singlet phase
for $J'/J <(J'/J)_c$.\cite{Matsumotoetal01} Hence, for $J'\ll J$ the
ground state is in a quantum disordered phase with a finite spin gap
$\Delta_0\approx J - zJ' + O(J'^2)$, where $z$ is the coordination
number for the dimers.  This gap can be overcome by a magnetic field
$h_c=\Delta_0$, at which a quantum phase transition to an
antiferromagnetically ordered phase occurs. This phase transition is
well described by the picture of Bose-Einstein condensation of
triplet quasiparticles.~\cite{Nikunietal00, Matsumotoetal04}

In a randomly doped system ($p >0$), the zero-field ground state
deviates strongly from a dimer singlet. As described in the introduction,
any arbitrarily small (but finite) site dilution with concentration
$p$ creates an interacting random network of LMs which
orders at $T=0$. \cite{Laflorencieetal04} When looking at
the physics of the system for energies much below
the gap $\Delta_0$, and consequently for fields $h\ll \Delta_0$,
the behavior of the diluted system Eq.~\eqref{e.ham_dilute} can be
captured by an effective spin-$1/2$ Heisenberg model for the
random network of LMs, whose Hamiltonian reads
\begin{equation}
H=\sum_{i<j}J^{\rm (eff)}_{ij}{\bm S}_i\cdot
{\bm S}_j-h\sum_i\emph{S}^z_i.
\label{e.ham_LRspin}
\end{equation}
Here the ${\bm S}$ operators represent $S=1/2$ LMs, exponentially
localized around the sites of unpaired spins. Hence, on a $L\times L$ lattice
the number of effective spins is approximately $N_s=pL^2$
(ignoring the case in which a dimer is fully replaced
by the impurities, and which does not produce any LM).
Since the diluted sites are randomly
located throughout the system in Eq.~\eqref{e.ham_dilute},
in the effective model Eq.~\eqref{e.ham_LRspin} the LMs
are also assumed to be randomly distributed on the square lattice.
The effective interactions $J^{\rm (eff)}_{ij}$ between LMs
$r_{ij}$ take the asymptotic form for large inter-moment distance
\cite{SigristF96,Mikeskaetal04,
Roscilde06}
\begin{equation}
J^{\rm(eff)}_{ij}=(-)^{|i-j|-1}\frac{J_1}{r_{ij}^\alpha}e^{-r_{ij}/\xi_0},
\label{e.J_LR}
\end{equation}
where $J_1$ is a parameter determined by $J$ and $J'$ of the
original diluted model. $\xi_0 \sim \Delta_0^{-1}$ is the
correlation length of the gapped ground state in the clean limit.
$J^{\rm(eff)}_{ij}$ has ferromagnetic /antiferromagnetic character depending
on whether the sites $i$ and $j$ belong to the same/different
sublattices; hence the couplings are not frustrated, and they favor
AFLRO of the LMs. $\alpha$ is a dimension-dependent exponent, taking
values $\alpha=0$ (1D), $\alpha=1$ (2D), and $\alpha=3/2$ (3D). In
the following we will concentrate on the 2D case. When $r_{ij}=1$ we
need to recover the limit $J^{\rm (eff)}_{ij}=J_1e^{-1/\xi_0} =J'$
for the direct interaction of two unpaired spins through a $J'$
coupling. This leads to the identification $J_1=J'e^{1/\xi_0}$,
which in turn implies
\begin{equation}
J^{\rm(eff)}_{ij}=(-)^{|i-j|-1}\frac{J'}{r_{ij}}e^{-(r_{ij}-1)/\xi_0}.
\label{e.J_LR1}
\end{equation}
Actually the second-order perturbation calculation
which delivers the asymptotic form of the effective
couplings, Eq.~\eqref{e.J_LR},
gives $J_1\sim (z J')^2/J$; this means that
the choice $J_1 \sim J'$, which correctly reproduces
the short-range behavior of the effective couplings,
leads to an overestimation of such couplings
in the long-distance limit. Such an overestimation
is nonetheless irrelevant, given the exponential
suppression of $J^{\rm(eff)}_{ij}$ at
large inter-moment distances.


In the following, we apply the Stochastic Series Expansion (SSE)
quantum Monte Carlo (QMC)~\cite{SyljuasenS02} to both the original and effective
Hamiltonians, given by Eq.~\ref{e.ham_dilute} and
~\ref{e.ham_LRspin}, in order to investigate the evolution of the ground
state of the system as a function of the applied magnetic field. For the
effective model with $N_s$ randomly distributed spins and
non-frustrated long-range exchange interactions, the SSE approach
has the advantage of requiring a computational
effort $O(N_s\ln N_s)$ instead of the naive $O(N_s^2)$.~\cite{Sandvik03}
To efficiently reach the ground-state behavior
via the finite-temperature SSE we make use of a $\beta$-doubling
approach~\cite{Sandvik02A} up to an inverse temperature
$\beta J=2^{15}$. The full statistics of the disorder
distribution is well reproduced by
averaging all results over $\approx 300$
disorder realizations.

In the presence of an applied magnetic field, spontaneous
AFLRO is reflected by the appearance of a finite
transverse staggered magnetization, estimated as
\begin{equation}
m_s=\sqrt{S^\bot(\pi,\pi)/L^2}, \label{e.Mstagger}
\end{equation}
where $L$ is the linear system size and $S^\bot(\pi,\pi)$ is the
transverse staggered structure factor, defined as
\begin{equation}
S^\bot(\pi,\pi)=\frac{1}{2L^2}{\sum_{ij}}(-)^{|i-j|}\langle
S_i^xS_j^x + S_i^yS_j^y\rangle.\label{e.ssf}
\end{equation}
The summation runs over all pairs of occupied sites in the original
site-diluted model, and it runs over all possible pairs of LMs in the
effective spin model. We also estimate the spin stiffness,
corresponding to a superfluid density upon spin-to-boson
mapping, via the fluctuations of
the winding numbers $W_{X,Y}$ of the SSE worldlines
along the two lattice directions $X$ and $Y$  \cite{PollockC87}:
\begin{equation}
\rho_s=\frac{1}{2\beta J}\langle W_X^2+W_Y^2\rangle. \label{e.Rho_s}
\end{equation}

\section{Monte Carlo Results for weakly coupled dimers}\label{ss.PD_weak}

In this section we present the QMC results for the full evolution
of the ground state
of site-diluted weakly coupled dimers
upon varying an applied magnetic field.
Here we focus on the case of weakly coupled dimers
by choosing $J'/J=1/4$. In the clean case, this system
is far away from the critical point $J'_c\approx
0.523J$ in zero field, and it exhibits a disordered
state with a correlation length
$\xi_0\approx 1$. When a magnetic field is applied, the clean
system experiences a quantum phase transition to a field-induced
antiferromagnetically ordered phase at
$h_{c1}^{(0)}=\Delta_0=0.60(1)J$. Doping with non-magnetic
impurities produces LMs. Our choice of $J' \ll \Delta_0$ guarantees
that the energy scale for the interactions between LMs
is well separated from the energy scale of interactions for
the intact dimers. Hence the application of a weak field $h\ll J$
to the system will exclusively probe the response of LMs,
while intact dimers remain essentially frozen in local
singlets.

\begin{figure}[h]
\begin{center}
\null~~~~~~\includegraphics[
clip,width=80mm,angle=0]{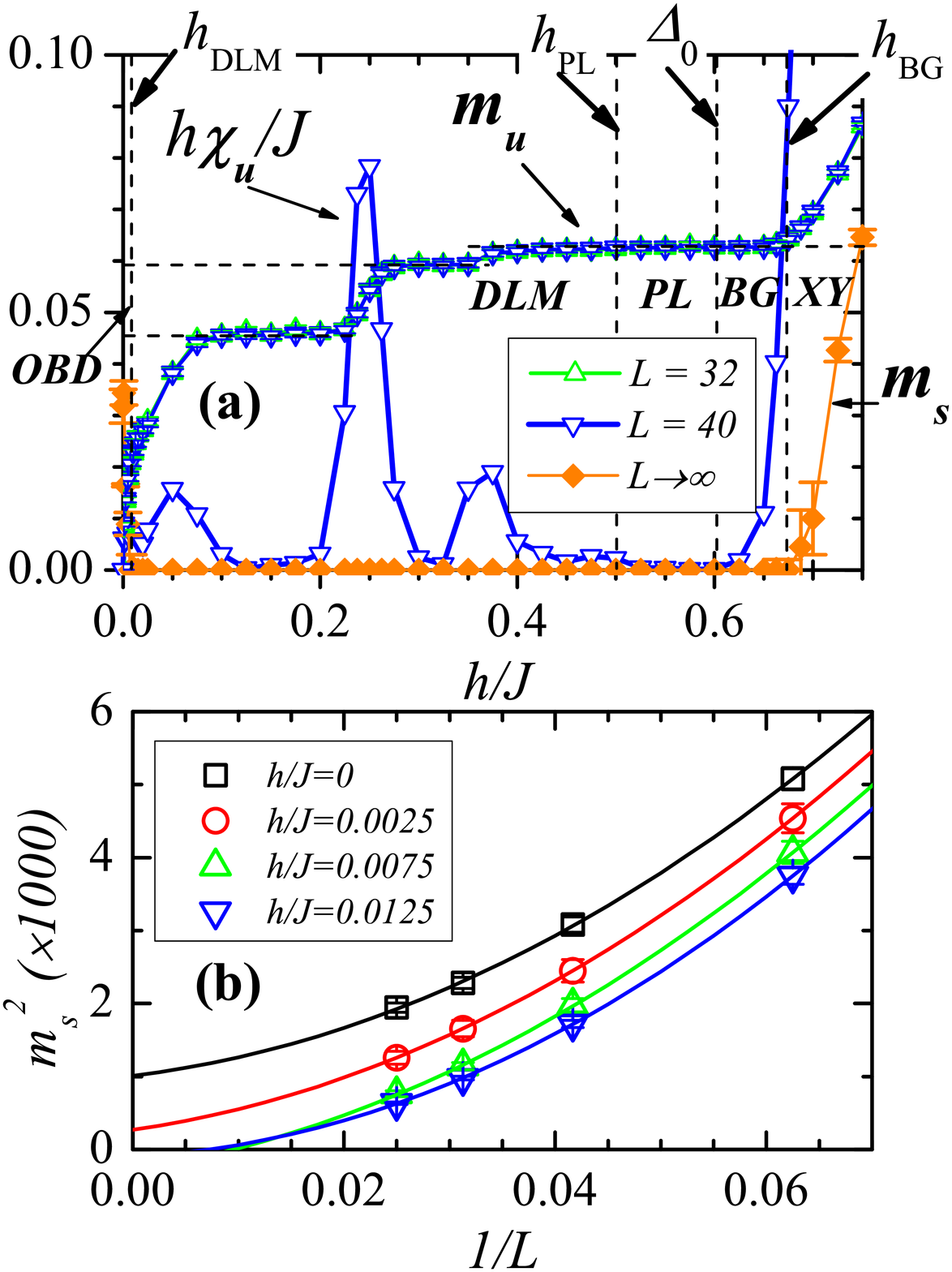}
\caption{(a): Field dependence of the uniform magnetization,
the uniform susceptibility, and the staggered magnetization,
of weakly coupled dimers. The coupling ratio is $J'/J=1/4$,
and the dilution concentration is $p=1/8$. The orange diamonds show the
staggered magnetization in the thermodynamic limit,
extrapolated by finite-size scaling as shown in (b). OBD:
order-by-disorder phase; DLM: disordered-local-moment phase; PL:
plateau phase; BG: Bose glass phase; and XY: XY-ordered
phase (magnetic condensate).} \label{f.FieldScan}
\end{center}
\end{figure}

 The succession of field-induced phases in the site-diluted
 system is fully captured by investigating the
 staggered magnetization, $m_s$ (Eq. \ref{e.Mstagger}),
and the uniform magnetization,
$m_u=\sum_i\langle S_i^z\rangle/L^2$. The former detects
the presence of order in the ground state, while the
latter probes the low-energy spectrum both in the ordered
and in the disordered phase, exhibiting the presence of
a triplet gap via magnetization plateaus.
The field dependence of these quantities
at a dilution concentration $p=1/8$ is
shown in Fig.~\ref{f.FieldScan}. The following succession
of phases is observed
in the field range $0\leq
h<J$: order-by-disorder phase (OBD),
disordered-local-moment phase (DLM), plateau phase (PL), Bose glass
phase (BG), XY-ordered phase (XY).

At zero field, a finite staggered magnetization
develops in the system due to the effective interactions
among LMs: an extrapolation of the QMC data to infinite
size gives $m_s= 0.032(3)$ (see Fig.~\ref{f.FieldScan}(b)),
indicating that the system is in the OBD phase.
Further careful finite-size scaling analysis, discussed in
Sec.~\ref{s.PD_entire}, shows that the OBD phase also survives
at small finite fields up to $h=h_{\rm DLM}=0.007(1)J\ll J'$.
At this critical field the {\it uniform} magnetization
takes the value $m_u=0.0208(6)$, namely it is
\emph{much less} than the value $m_u^{\rm sat}=pS=1/16$ corresponding
to full saturation of the local moments. In fact, the saturation
value is only attained at a much higher field $h=h_{\rm PL}\approx 2J'$.
Hence the ground state of the diluted system at $h_{\rm DLM}<h<h_{\rm PL}$
has a striking feature: the LMs have lost spontaneous ordering
transverse to the field, but they are far from being all polarized.
In contrast, in a clean antiferromagnet the destruction of
a spontaneously ordered state is \emph{always} accompanied by the full
polarization of its constituents. Moreover for $h>h_{\rm DLM}$
the magnetization keeps growing with the field, revealing a
gapless spectrum in the absence of spontaneous ordering.

For $h_{\rm PL}<h<\Delta_0$ the uniform magnetization
remains constant at $m_u^{\rm sat}$, indicating a gapped,
disordered plateau phase~ (PL)\cite{Mikeskaetal04} in which
the field completely aligns all the unpaired spins, but is not
sufficiently strong to break the singlets on strongly coupled dimers.
The existence of the PL phase is then a characteristic of the
weakly coupled dimer system, and we will see that it disappears
when $J'\sim \Delta_0$.

When $h\geqslant\Delta_0$ the gap to excite a triplet in a region of
intact dimers (\emph{dimer triplet}) closes just as it does in the clean
system. The reason for this is the finite (albeit exponentially
small) probability of finding an arbitrarily large clean region
which is devoid of any impurity; the local gap of such a clean
region can be arbitrarily close to $\Delta_0$ so that a field
$h\gtrsim\Delta_0$ can create a localized dimer triplet in that
region. At variance with the clean system, the field-induced
localized quasiparticles do not condense in an extended state with
finite superfluid response (spin stiffness), but they are instead
localized by disorder and give rise to a quantum disordered Bose glass
state.~\cite{RoscildeH05,Roscilde06} In fact the fully
polarized LMs act essentially as impenetrable barriers to dimer
triplets, while intact dimers close to the non-magnetic impurities
have higher local gaps to dimer triplets than in the clean regions.
Hence site dilution creates an effective disorder potential which
has the effect of Anderson-localizing the dimer triplet
quasiparticles appearing in the ground state.~\cite{Fisheretal89}

The resulting BG phase extends over the field range $0.60(1)
\leqslant h/J\leqslant 0.67(1)$ in Fig.~\ref{f.FieldScan}(a), and,
similarly to the DLM phase, it is identified by the absence of a
gap, as shown by the finite slope in the magnetization curve (namely
a finite susceptibility $\chi_u = \partial m_u / \partial h$).
\cite{RoscildeH05, RoscildeH06, Roscilde06}

At $h_{\rm BG}/J\approx0.67$, the applied magnetic field finally
drives the system to an XY-ordered antiferromagnetic state. In the
picture of bosonic triplet quasiparticles, this magnetically ordered
state is a condensate with finite superfluid response. Hence this
system realizes a magnetic BG-SF transition, with critical
exponents which are quite different from the more conventional Mott
insulator to superfluid (MI-SF) transition. \cite{RoscildeH05, 
RoscildeH06, Roscilde06}
As $h$ is increased
to reach half filling of the dimer triplets, the behavior of the
system is better understood in terms of singlet quasiholes (given
that the triplet quasiparticles are effectively hardcore). Such
quasiholes experience the reverse succession of phases to that of
triplet quasiparticles, namely a transition from XY to BG and then
from BG to a plateau phase corresponding to full polarization of the
entire sample.~\cite{RoscildeH05}

\section{Field-Induced DLM Phase}\label{s.QDPhase}

\subsection{Magnetization curve and effective model}

In this section we focus
on the unconventional properties of the DLM phase.
As already pointed out in the previous section,
this phase is characterized by the absence of
a gap, as revealed by a finite susceptibility.
This is actually not at all obvious from Fig.~\ref{f.FieldScan},
where the susceptibility appears to nearly vanish
in correspondence with intermediate \emph{pseudo-plateaus} (PP)
at $73\%$ and $95\%$ of the
saturation magnetization. Only a careful analysis
of the temperature scaling of the susceptibility
\cite{Yuetal08} shows
that this quantity remains finite at all field values
in the DLM phase down to $T=0$. The first PP extends up
to $h\approx 0.7J'$; the second appears at $h\approx 1.2J'$;
and the true saturation plateau is only attained at $h\approx 2J'$.

\begin{figure}[h]
\begin{center}
\null~~~~~~\includegraphics[
bbllx=30pt,bblly=10pt,bburx=500pt,bbury=610pt,%
clip,width=80mm,angle=0]{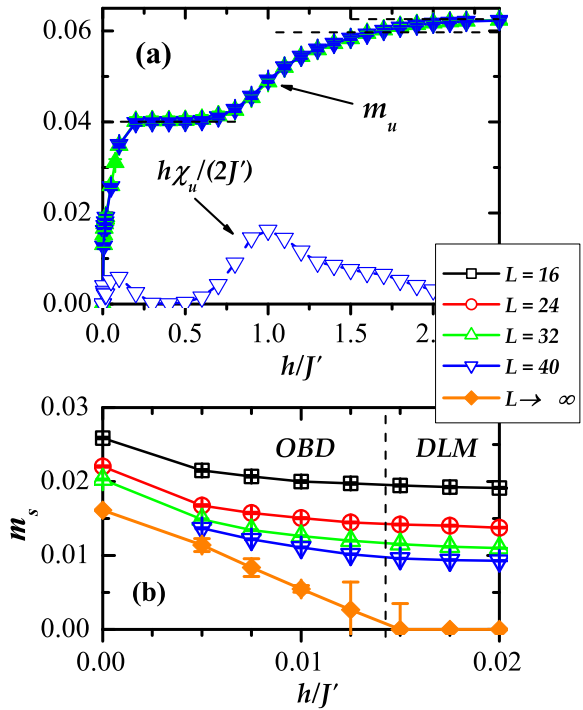}
\caption{Field dependence of $m_s$, $m_u$, and $\chi_u$ in the
effective model, Eq.~\eqref{e.ham_LRspin}, with $\xi_0=1.0$ at a dilution concentration
$p=1/8$.} \label{f.FieldScanLR}
\end{center}
\end{figure}

As argued in Sec.~\ref{s.Models} the physics of the site-diluted
dimer model at low energy (namely at low fields for $T=0$)
can be fully captured by an effective model of coupled
LM, Eq.~\eqref{e.ham_LRspin}. Here we make this statement more
quantitative, by directly investigating the physics
of the effective model via QMC simulations. The field
evolution of the uniform and staggered magnetization
for the model of Eqs.~\eqref{e.ham_LRspin} and \eqref{e.J_LR} with
$\xi_0 = 1$ is shown in Fig.~\ref{f.FieldScanLR}.
Similarly to the results for the
original diluted model, we observe the destruction of
the AFLRO phase of the LMs at very small fields, and the
occurrence of an extended DLM phase. The
occurrence of PPs in the magnetization curve is
fully captured in the effective model, albeit with
less sharp features. The full qualitative correspondence
between the original Hamiltonian and the effective LM model
corroborates the picture that the magnetic response
of the system up to the plateau phase is dominated
by LMs. In the following subsection we will bring
this observation to a microscopic level,
relating the occurrence of PPs to the quantum magnetic
response of LM clusters.


\begin{figure}[h]
\begin{center}
\null~~~~~~\includegraphics[
clip,width=80mm,angle=0]{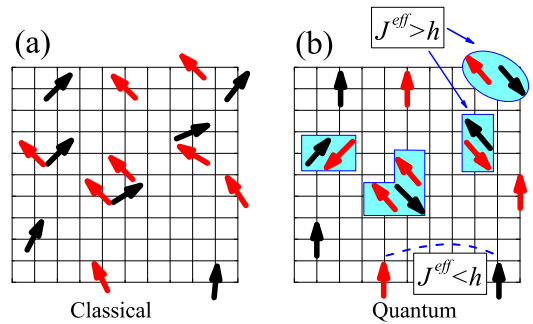}
\caption{(a): Classical $S\rightarrow\infty$ local moments form a
canted antiferromagnetic order at finite field. (b): In the quantum
case, neighboring local moments prefer to form strongly
correlated states on local clusters,
leading to the novel DLM phase. The black and red arrows distinguish
spins located on different sublattices.} \label{f.DFMscenarios}
\end{center}
\end{figure}

\begin{figure}[h]
\begin{center}
\includegraphics[
bbllx=20pt,bblly=20pt,bburx=750pt,bbury=580pt,%
width=80mm,angle=0]{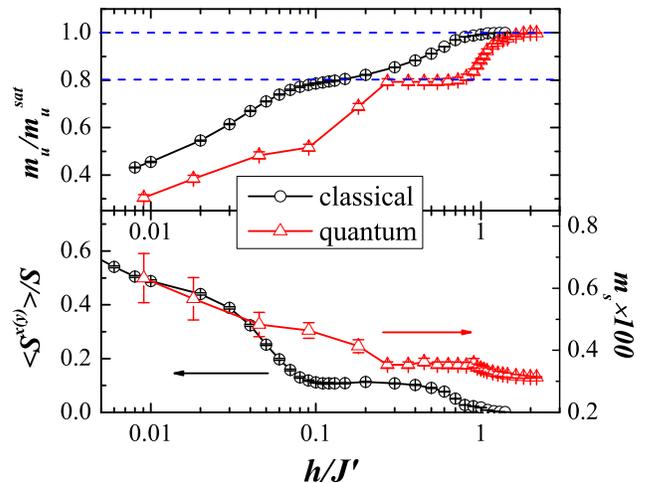}
\caption{Field dependence of $m_u$ and of the transverse spin
component $S^{x(y)}$ in the classical effective spin model with
$\xi_0=1.0$, $p=0.0625$ and $L=40$. The corresponding $m_u$ and
$m_s$ of the quantum model with the same model parameters are also
shown for comparison. The staggered magnetization $m_s$ of the
quantum model has been magnified by a factor of
$100$.}\label{f.MuSxClassical}
\end{center}
\end{figure}

\subsection{Quantum nature of the disordered-local-moment phase}
\label{ss.clusterscenario}

 In order to emphasize the genuinely quantum character
 of the DLM phase, in this subsection we focus on a
 comparison of the system of quantum
 $S=1/2$ LMs described by Eq.~\eqref{e.ham_LRspin}
 with its classical limit $S=\infty$.
The classical limit of Eq.~\eqref{e.ham_LRspin}
is investigated via a
classical Monte Carlo study of a $40\times 40$ lattice
with a concentration $p=1/16$ of LMs. The ground state
of the classical system is reached by careful
thermal annealing with a linear protocol, and results are averaged over
typically $500-1000$ configurations.

In the classical limit at $T=0$, the system of randomly distributed LMs
in a sufficiently weak magnetic field exhibits a canted antiferromagnetic
ground state (Fig.~\ref{f.DFMscenarios}(a)).
Upon increasing the field,
the most isolated LMs minimize their energy by fully aligning with
the field, but the remaining LMs preserve long-range order
transverse to the field because of the direct
long-range effective interactions,
as clearly exhibited by Fig.~\ref{f.MuSxClassical}. Long-range order
is destroyed only when the field exceeds the polarization 
value of the largest connected LM cluster that can form in the system, 
thereby polarizing all the LMs.
This means that, in the classical limit,
the destruction of AFLRO is always accompanied by the
full saturation of the magnetization.

Fig.~\ref{f.MuSxClassical} compares the classical
behavior with the one of the quantum system with the
same concentration of LMs. 
The uniform magnetization curve shows the emergence of
a PP also at the classical level, precisely at the
same value as in the quantum case. This result
is consistent with the picture (both classical
and quantum) of PPs originating
from strongly coupled LMs which resist to being
fully polarized by the field.
This picture relies on the strong inhomogeneity
of the network of interacting LMs. This network
exhibits an average distance between LMs of
$ r_{\rm ave}  = 1/\sqrt{p}$, and
a corresponding average coupling
$J_{\rm ave} \ll J'$ in the limit of diluted
LMs. Yet the magnetic response is strongly
dominated by the long tails of
distribution of LM couplings and \emph{not} by
its average. In particular, there exist
small clusters of LMs with linear size $r\ll r_{\rm ave}$,
which interact with much stronger couplings
$J^{\rm(eff)}\sim J'$.
The main examples of such clusters are
nearest-neighboring dimers and trimers
(see the cartoon of Fig.~\ref{f.DFMscenarios}(a)).
Evidently only a field $h \sim J'$ can overcome the
antiferromagnetic correlations developing in these clusters.

 In the classical limit, LM clusters not only retain
short-range antiferromagnetic correlations due to their
strong internal coupling, but also long-range
correlations due to the direct long-range couplings
existing among them.
Hence, AFLRO classically survives the strong
field $h\sim J'$, as shown in Fig.~\ref{f.MuSxClassical}.
In particular the transverse staggered
magnetization shows a PP corresponding to
that of the uniform magnetization, a fact which
corroborates the above picture.

 On the contrary, the quantum mechanical
behavior of the LM network is significantly different,
as already shown by the data on transverse staggered
magnetization in Fig.~\ref{f.FieldScan}. At the quantum level
the presence of strong short-range antiferromagnetic
correlations  persisting on LM clusters does \emph{not}
coexist with long-range antiferromagnetic correlations
between different clusters.  In fact a
phenomenon of quantum \emph{clustering} of
antiferromagnetic correlations (dimerization,
trimerization \emph{etc.}) turns out to minimize
the energy at the quantum level, as sketched
in Fig.~\ref{f.DFMscenarios}(b). As it will be quantitatively
shown in the next subsection, LM dimers have the tendency
to freeze in a singlet, LM trimers tend to freeze in a
$S_{\rm tot}=1/2$ state aligned with the field, etc.
LM dimer singlets are obviously uncorrelated with
one another, while LM trimers are also antiferromagnetically
uncorrelated because the $S_{\rm tot}=1/2$ state does
not reflect the same sublattice structure for all trimers.
Hence, except for a small field range slightly above
zero field, the entire phase with partially polarized
LMs is quantum disordered.

\begin{figure}[h]
\begin{center}
\null~~~~~~\includegraphics[
     width=80mm,angle=0]{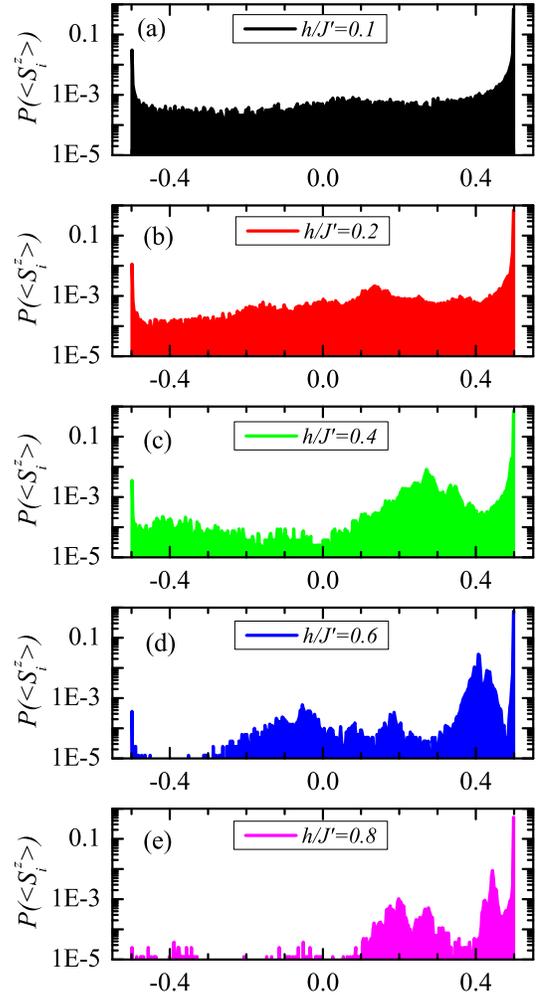}
\caption{Distribution of the $z$ component of local moments $S_i^z$
in the effective spin model for classical spins with $\xi_0=1.0$,
$p=0.0625$, and $L=40$.} \label{f.Sz_classical}
\end{center}
\end{figure}

\subsection{Local Moment Clusters}\label{ss.Distribution}

The dominant role played by the 
strongly interacting clusters of LMs in the DLM phase
clearly emerges from the distribution
of the local magnetization $\langle S_i^z \rangle$,
as well as of the pair correlations
$C(r_{ij}) = \langle {\bm S}_i\cdot {\bm S}_j \rangle$,
developed by the inhomogeneous network of LMs.
The quantum behavior of these
small LM clusters is revealed again by comparison
with the classical limit.

  The distribution of $\langle S^z_i \rangle$
in the classical model is shown in Fig.~\ref{f.Sz_classical}.
Here the classical spin length is normalized to 1/2
to compare with the quantum case.
At zero field, $\langle S^z_i\rangle$ is obtained by averaging the
ordered ground state over all possible orientations.
The corresponding probability distribution is:
\begin{equation}
P(\langle S^z_i \rangle) = \frac{1}{\sqrt{1-\langle S^z_i \rangle^2}}
\end{equation}
The presence of a very weak field further
enhances 
two peaks located at $\langle
S^z_i\rangle=\pm0.5$. The peak at $\langle S^z_i\rangle=0.5$
corresponds to spins fully polarized by the magnetic field, whereas
the peak at $\langle S^z_i\rangle=-0.5$ corresponds to spins
antiferromagnetically coupled to the fully polarized ones (not shown
in Fig.~\ref{f.Sz_classical}). For
$h/J'>0.1$, an extra peak appears at a positive
$\langle S^z_i\rangle$, and it moves towards
$\langle S^z_i\rangle=0.5$ with increasing magnetic field:
this peak corresponds to the partially polarized (canted)
LMs. At the same time, the peak at $\langle S^z_i\rangle=-0.5$ drops.
This is fully consistent with the picture that classical spins
are continuously polarized.


\begin{figure}[h]
\begin{center}
\null~~~~~~\includegraphics[
bbllx=20pt,bblly=20pt,bburx=370pt,bbury=620pt,%
     width=80mm,angle=0]{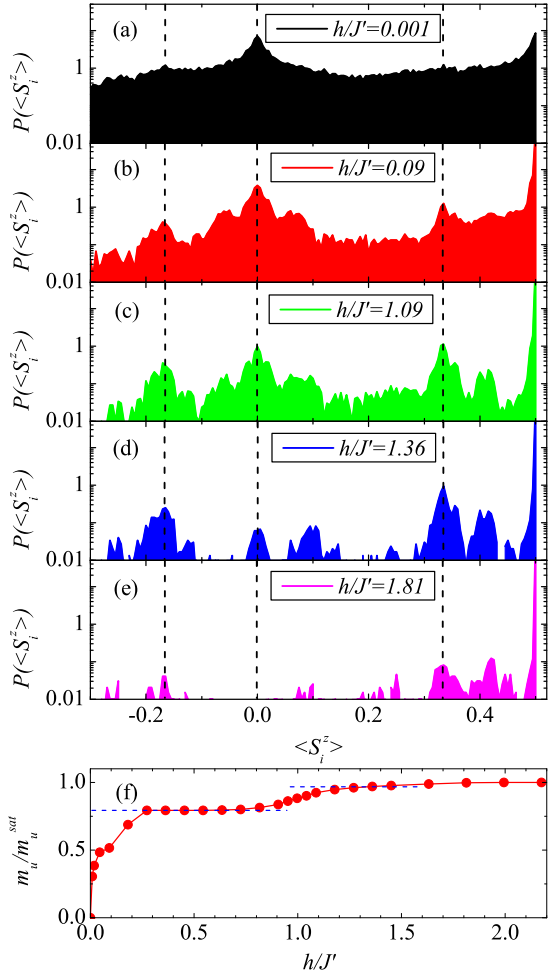}
\caption{(a)-(e): Distribution of the $z$ component of the local moments
$S_i^z$ in the effective spin model for quantum spins in various
fields with $\xi_0=1.0$, $p=0.0625$, and $L=40$; (f): Field
dependence of the uniform magnetization $m_u$ for the same model
parameters.} \label{f.Sz_qLRh}
\end{center}
\end{figure}

 The local magnetization histograms for the quantum system
of LMs are shown in Fig.~\ref{f.Sz_qLRh}, and they
exhibit significant differences with respect to the classical
case, related to the quantum behavior of LM clusters.
For $h=0$ (not shown in Fig.~\ref{f.Sz_qLRh}) the histogram for $\langle S^z_i \rangle$
exhibits a unique $\delta$-peak at $\langle S^z_i \rangle = 0$,
which obviously corresponds to the fact that the
ground state of a quantum Heisenberg antiferromagnet on a
finite-size system is a total singlet.
The evolution of the local-magnetization
histograms is strongly non-monotonic upon applying an increasing
field, and it allows to precisely track the succession
of phases in the LM network. In fact for a small field,
corresponding to the OBD phase, the
$\langle S^z_i \rangle=0$ peak evolves into a
broad distribution of local magnetizations,
which gives a direct insight into the microscopic
structure of the OBD state. A significant
portion of the LMs is in a canted antiferromagnetic
phase with variable canting, depending on the
local structure of LM couplings. This
corresponds to the tail of the distribution
at $\langle S^z_i \rangle >0$. At the same
time a significant portion of the LMs is
fully polarized by the field, as shown by the
emergence of a peak at $\langle S^z_i \rangle=1/2$.
These polarized LMs create an effective static field for the LMs which
are coupled to them. In the case of antiferromagnetic couplings,
this field points opposite to the external field, and it induces
a fraction of the LMs to develop a negative
local magnetization, as shown by the fat tail
of the histogram at $\langle S^z_i \rangle < 0$.
Yet the persistence of a central peak
at $\langle S^z_i \rangle = 0$ shows that
a large fraction of the LMs remains \emph{frozen}
in \emph{local singlets}, associated
with even-number LM clusters (dimers, quadrumers,
etc.). At the same time two further peaks
clearly emerge in the histogram, at
 $\langle S^z_i \rangle = 1/3$ and at
 $\langle S^z_i \rangle = -1/6$: they can
 be easily associated with the state
$|S_{\rm trimer} = 1/2,
 S^z = 1/2 \rangle$ for LM trimers, where
 one has a trimer magnetization
 $S^z = \sum_{i \in {\rm trimer}} \langle S^z_i \rangle =
 1/3 + 1/3 - 1/6 = 1/2$. 
 Further (and smaller) peaks are to be attributed to larger
 (and more rare) LM clusters. 

  The singlet peak at $\langle S^z_i \rangle = 0$
 persists over a significant field range. In particular
 in the field interval $0.3 \lesssim h/J' \lesssim 0.8$
 the local-magnetization histogram changes very little, corresponding to
  the appearance of the first PPs
  (as presented in Fig.~\ref{f.Sz_qLRh}(f)). Only for
  $h>J'$, namely when the field overcomes the largest
  possible triplet gap associated with a LM dimer-singlet,
  we observe a decrease of the singlet peak and correspondingly
  an increase in the slope of the magnetization curve,
  which allows then to associate the first PP primarily with the
  quantum magnetic response of LM dimers. Remarkably
  the LM-trimer peaks at $\langle S^z_i \rangle = 1/3$
  and $-1/6$ persist due to the larger gap to full
  saturation exhibited by these structure. The weak 
  field dependence of such peaks for fields
  in the interval $1.2 \lesssim h/J' \lesssim 1.4$
  allows then to unambiguously associate the second PP
  to the LM trimer gap from $|S_{\rm trimer} = 1/2,
 S^z = 1/2 \rangle$ to $|S_{\rm trimer} = 3/2,
 S^z = 3/2 \rangle$, corrected by the
 local field interacting with each trimer in the network.

\begin{figure}[h]
\begin{center}
\null~~~~~~\includegraphics[
clip,width=80mm,angle=0]{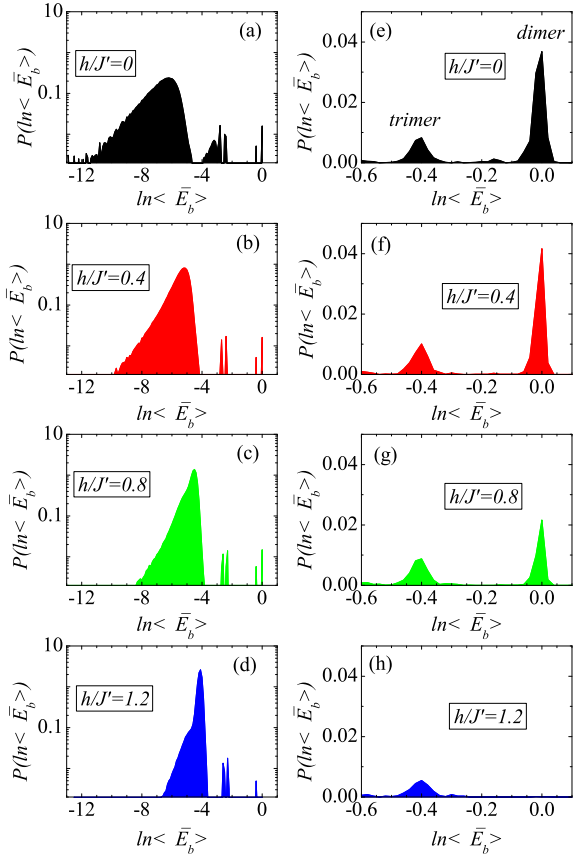}
\caption{(a)-(d): Distribution of the normalized bond energies
$\langle\bar{E_b}\rangle$ in the effective spin model with
$\xi_0=1.0$, $p=0.0625$ and system size $L=40$. (e)-(h): 
Zoom on the distribution in the region exhibiting the dimer and
trimer peaks.} \label{f.Eb_LR}
\end{center}
\end{figure}

The correspondence of the PP to the magnetic
response of LM clusters is further
confirmed by investigating
the distribution of bond energies.
The minimum bond energy is the one of the singlet
$E_b=-\frac{3}{4}J'$. We hence define a normalized bond energy
$\bar{E}_b=-4E_b/(3J')$, upper-bounded by 1,
and we plot the field dependence of the
distribution of $\ln\langle\bar{E}_b\rangle$ in Fig.~\ref{f.Eb_LR}.
In zero field we observe a very broad distribution,
characteristic of a highly inhomogeneous system.
In particular there is a very broad peak at
low energies (in absolute value), corresponding to
the long tail of the distribution of LM couplings. But the
most striking feature is a gap between the low-energy
peak and two peaks corresponding to the stronger
bonds, $\langle\bar{E}_b\rangle=1$ and
$\langle\bar{E}_b\rangle=2/3$. These peaks clearly
correspond to dimer singlets and to trimer doublets,
respectively. They are very resistant
to the application of a field, while the low-energy
part of the distribution changes continuously,
developing a strong peak which corresponds
to the LMs fully polarized by the field. 
The dimer peak starts to decrease
continuously for $h \gtrsim 0.4 J'$, and it disappears
for $h \gtrsim J$, corresponding to the end
of the first PP. This in turn reveals that the
triplet gaps of the LM dimers follow a distribution,
upper bounded by $J'$, and
due to the effect of the local effective field exerted
by the other LMs surrounding the dimer and
randomly coupled to it. Therefore the ensemble of
LM dimers responds to a field in a continuous
fashion, which gives rise to the finite slope
of the first PP. Similarly, the trimer peak starts
to decrease only for $h \gtrsim 1.2$, consistent
with the appearance of the second PP with
finite slope.

\subsection{Long-range clusters}

 A fundamental result in one-dimensional quantum antiferromagnets with
random nearest-neighbor couplings is the formation of spin singlets
at all energy scales and all distances, to form the so-called
random-singlet phase.~\cite{Fisher94} This phase is fully captured
by a real-space renormalization group approach which, at each step,
identifies the locally strongest antiferromagnetic bonds and freezes
the corresponding spins into a singlet, obtaining then effective
couplings for the leftover spins via second-order perturbation
theory around the singlet state. The application of a similar
approach to the two-dimensional case of the Heisenberg
antiferromagnet with random bonds fails to find a random-singlet
phase,~\cite{Linetal03} in agreement with quantum Monte Carlo
showing the persistence of N\'eel order.~\cite{Laflorencieetal06}

\begin{figure}[h]
\begin{center}
\null~~~~~~\includegraphics[
bbllx=40pt,bblly=20pt,bburx=500pt,bbury=670pt,%
     width=80mm,angle=0]{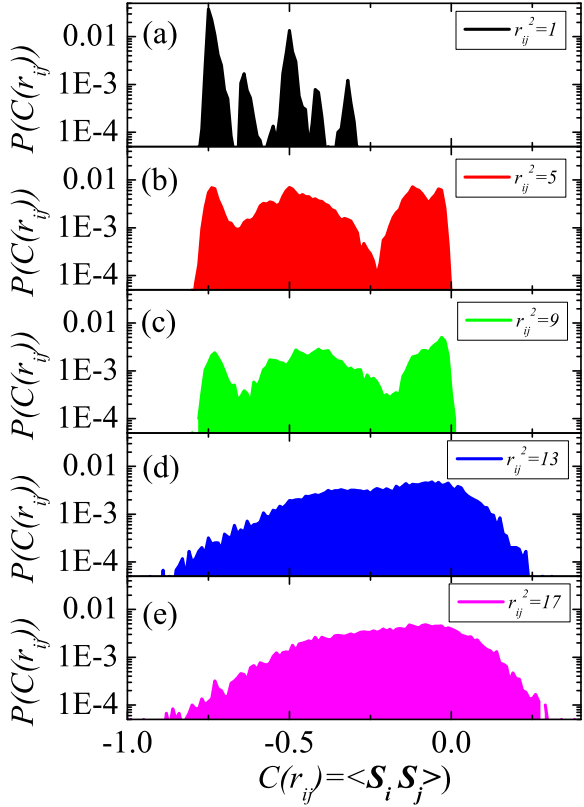}
\caption{Distribution of inter-moment correlation functions
$C(r_{ij})$ for various inter-moment distances $r_{ij}$ in the
effective model with $\xi_0=1.0$, $p=0.0625$, and at zero field. Here
sites $i$ and $j$ are always located on opposite sublattices.}
\label{f.CFLR}
\end{center}
\end{figure}

The effective model of interacting LMs, Eq.~\eqref{e.ham_LRspin},
shares similar features to those of a random-bond Heisenberg model,
namely the presence of a broad distribution of energy scales for the
LM couplings. As shown in this study, and as discussed extensively
in Ref.~\onlinecite{Laflorencieetal04}, the system of LMs in zero
field orders antiferromagnetically. We probe directly the possible
formation of longer-range singlets by investigating the distribution
of the correlation function $C(r_{ij})=\langle {\bm S}_{i}\cdot{\bm S}_{j}
\rangle $
for antiferromagnetically coupled LMs at various inter-moment
distances $r_{ij}$.
 If sites $i$ and $j$ form an approximate
singlet, then $C(r_{ij}) \approx -3/4$, while if they participate
in a long-range trimer, $C(r_{ij})
\approx -1/2$. The results for
$r_{ij}=1,\sqrt{5},3,\sqrt{13},\sqrt{17}$ in a system with
$p=1/16$ are shown in
Fig.~\ref{f.CFLR}. For $r_{ij}=1$, \emph{i.e.} for nearest-neighboring
moments, there are two sharp peaks in $P(C(r_{ij}))$ around
$C(r_{ij})=-3/4$ and $C(r_{ij})=-1/2$ respectively. As discussed
above, these two peaks correspond to LMs which participate in
nearest-neighboring dimers and trimers. A similar peak structure is
also observed for $r_{ij}=\sqrt{5}$ and $r_{ij}=3$, indicating
the appearance of similar structures at a larger length scale.
The only difference is that the dimer and trimer peaks are weaker
and broader, and that there is an extra peak around $C(r_{ij})=0$
for $r_{ij}=\sqrt{5}$ and $r_{ij}=3$, which corresponds to pairs
of spins belonging to \emph{uncorrelated} LM clusters (\emph{e.g.} two
short-range LM dimers or LM trimers).
 For $r_{ij}>3$ the distribution function completely changes its
shape. There is a very broad peak around $C(r_{ij})=0$ and a
shoulder around $C(r_{ij})\approx -1/3$, corresponding to long-range AF
correlations, but no peaks at $-3/4$ or $-1/2$.
It is interesting to notice that the average inter-moment
distance $r_{\rm ave}=1/\sqrt{p}=4$ roughly marks the
boundary of two behaviors. Hence the above numerical
findings imply that at zero field, LMs with inter-moment distance
$r_{ij}\lesssim r_{\rm ave}=1/\sqrt{p}$ are generally
involved in local singlets (doublets) associated with
even (odd) clusters local, and they only marginally take part in the AFLRO.
On the contrary the long-range AF correlations are carried by LMs with
$r_{ij}\gtrsim r_{\rm ave}$.
We finally notice that $J^{\rm(eff)}(r_{ij}=\sqrt{5})\sim h_{\rm DLM}$. Thus, in the
DLM phase almost all LMs involved in long-range clusters are fully polarized by the field,
and thus the magnetic properties of DLM phase mostly depend on the statistics of
the clusters made of nearest-neighboring LMs.

\subsection{Statistics of Local Clusters and Pseudo-Plateaus: \\a Minimal Model}
\label{ss.Magnetization}

 As seen in the analysis of the QMC data presented in the
 previous section, the physics of the
 DLM phase can be mainly ascribed to the behavior of
 clusters of nearest-neighboring LMs
 immersed in the local field which is created by the surrounding
 isolated LMs. In this section we show how to capture
 some fundamental aspects of the DLM phase with an elementary
 model inspired by the above results.

  The fundamental assumption we make is the diluteness
 of the LMs, namely $p\ll 1$. Under this assumption the
 majority of LMs do \emph{not} have nearest neighbors,
 and are typically at a distance $1/\sqrt{p}$ from the
 other LMs. We will denote these LMs as \emph{isolated} LMs.
 Assuming to have an applied magnetic field which is able to
 polarize most of the isolated LMs, we can essentially
 neglect the fluctuations of their magnetizations, and
 treat them as static variables. The remaining minority
 of LMs is arranged in clusters (dimers, trimers, etc.)
 which are rare and hence they seldom happen to be
 close to another. It is therefore a reasonable approximation
 to neglect all interactions between such clusters.
 Given that the LMs belonging to the clusters are the only dynamical
 variables left after freezing the isolated LMs,
 we can then model the magnetic response of the LM
 network in the DLM phase as the response of independent
 LM clusters immersed in the joint external field and
 in the local field created by the surrounding
 isolated LMs. For each cluster $C$ we therefore need
 to diagonalize the following effective Hamiltonian

 \begin{equation}
 {\cal H}_C = \sum_{i,j\in C} J_{ij}^{\rm (eff)}
 {\bm S}_i \cdot {\bm S}_j -
 \sum_{i\in C} \left( h - \sum_{j\in \bar{C}}
 \frac{J_{ij}^{\rm (eff)}}{2} \right) S^z_i
 \end{equation}
where $\bar{C}$ represents the complement to the
 cluster $C$. In this simplified model we have
 assumed that all spins outside the clusters
 are fully polarized, $\langle S^z_{j\in \bar{C}}\rangle = 1/2$.
 This is of course not exact due to the existence of
 other LM clusters, but if they are sufficiently far
 apart from the cluster under consideration (which is
 true if $p\ll 1$) we can neglect their couplings
 $J_{ij}^{\rm (eff)}$ to the cluster $C$, given the exponential
 decay of $J_{ij}^{\rm (eff)}$ with the distance.

 After diagonalizing ${\cal H}_C$ for each cluster, we may calculate
 the uniform magnetization via

 \begin{equation}\label{e.mu_cluster}
 m_u =\sum_{\{C\}}m_u(C)P(C),
 \end{equation}
 where $P(C)$ is the probability cluster $C$ appears. Since for each
 finite cluster $m_u(C)$ is a multi-step function at low temperature, $m_u$ is
 expected to exhibit a multi-step character modulated by the
 distribution $P(C)$.

\begin{figure}[h]
\begin{center}
\null~~~~~~\includegraphics[
bbllx=40pt,bblly=30pt,bburx=445pt,bbury=300pt,%
     width=80mm,angle=0]{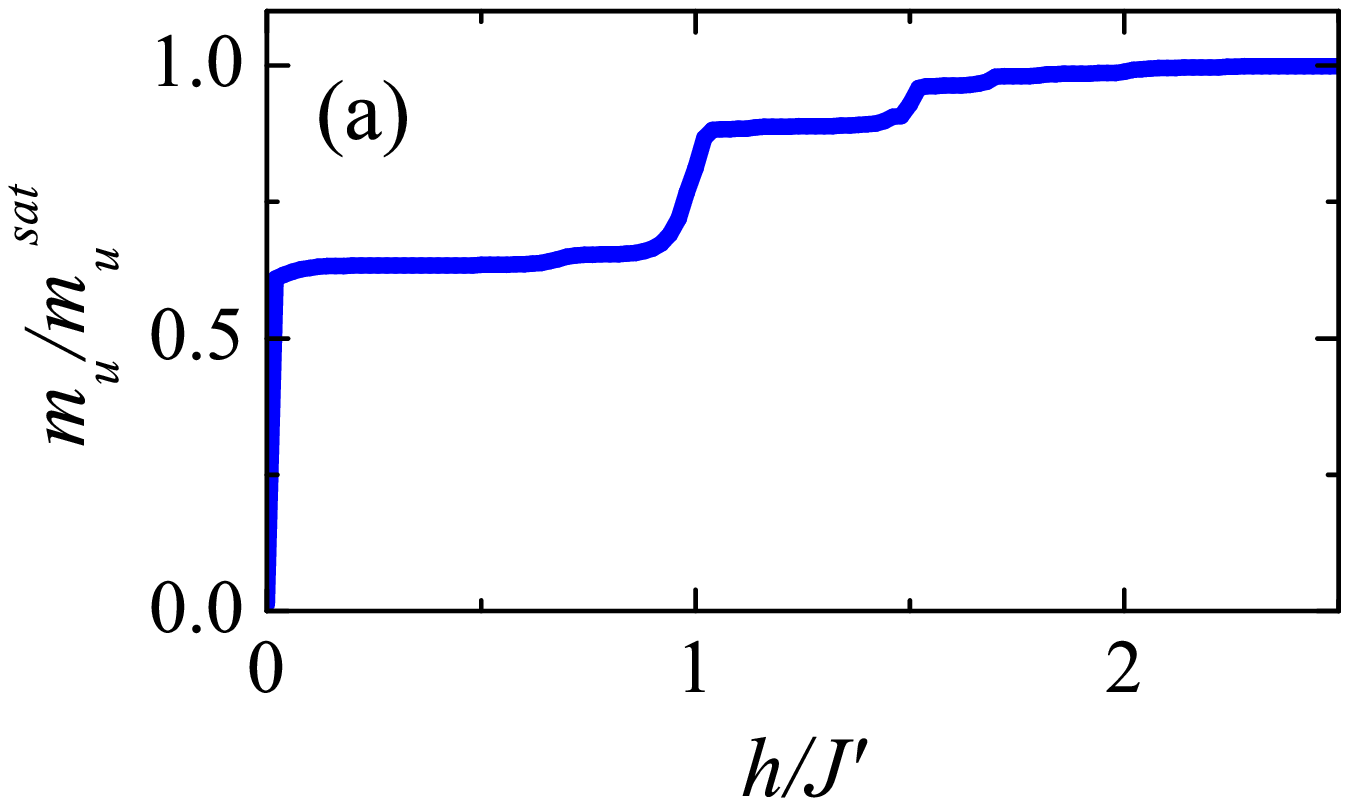}
\null~~~~~~\includegraphics[
bbllx=60pt,bblly=30pt,bburx=590pt,bbury=480pt,%
     width=80mm,angle=0]{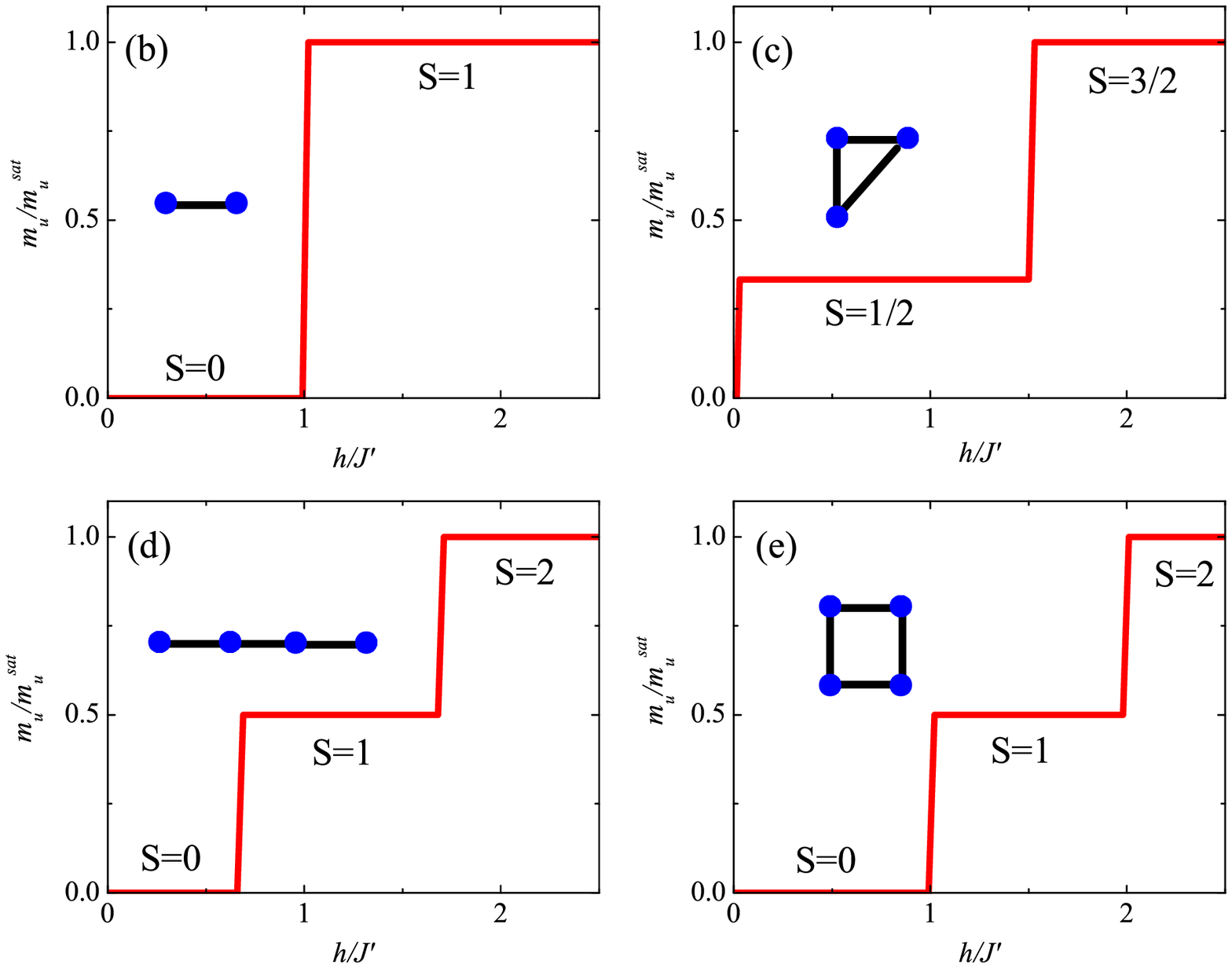}
\caption{Field dependence of the normalized magnetization
$m_u/m_u^{sat}$ calculated by applying Eq.~\ref{e.mu_cluster} in the
effective model with $\xi_0=1.0$, $p=1/8$, and $L=100$ shown in (a).
The normalized magnetization on small clusters up to $N=4$ with the
same model parameters are also shown in (b)-(e).} \label{f.Mbar}
\end{center}
\end{figure}

 In Fig.~\ref{f.Mbar}(a) the magnetization curve for the
 effective model with $\xi=1.0$, $p=1/8$, and $L=100$ is
 calculated by applying Eq.~\ref{e.mu_cluster}. We see that $m_u$ reaches
 the saturated value $m_u^{sat}$ only when $h/J'\gtrsim 2$. For $h/J'\leqslant2$,
 the curve resembles the one obtained by QMC, i.e., several PPs can be resolved.
 By comparing with the magnetization curves of small
 clusters presented in Fig.~\ref{f.Mbar}(b)-(e), we can identify the
 first PP at $h/J'\lesssim 0.7$ to be associated with the $S=0$ singlet
 states of even-number clusters and the $S=1/2$ states of odd-number clusters. Since $P(C)$ decays exponentially
 with the cluster size, this PP mostly comes from the $S=0$ state of LM
 dimers. Similarly we identify that the PP from $1.0 \lesssim h/J'\lesssim
 1.5$ is associated with the $S=1/2$ state of LM trimers. These are
 consistent with results obtained in previous sections.
 Interestingly, we can resolve other weak structures in the magnetic
 curve. For example, the little kinks at $h/J'\approx0.7$ and $h/J'\approx1.7$
 are associated with
 the $S=0$ to $S=1$ and $S=1$ to $S=2$ transitions of four-site
 chains. We also notice that $m_u$ saturates for $h/J'\geqslant2$ for small clusters
 at least with sizes up to $N=4$. Actually, $N>4$ LM clusters
 are extremely rare so that they have only negligible contribution
 to the magnetization curve. Hence we may take $h=2J'$ as the
 saturation field.

\begin{figure}[h]
\begin{center}
\null~~~~~~\includegraphics[
bbllx=30pt,bblly=20pt,bburx=450pt,bbury=540pt,%
     width=80mm,angle=0]{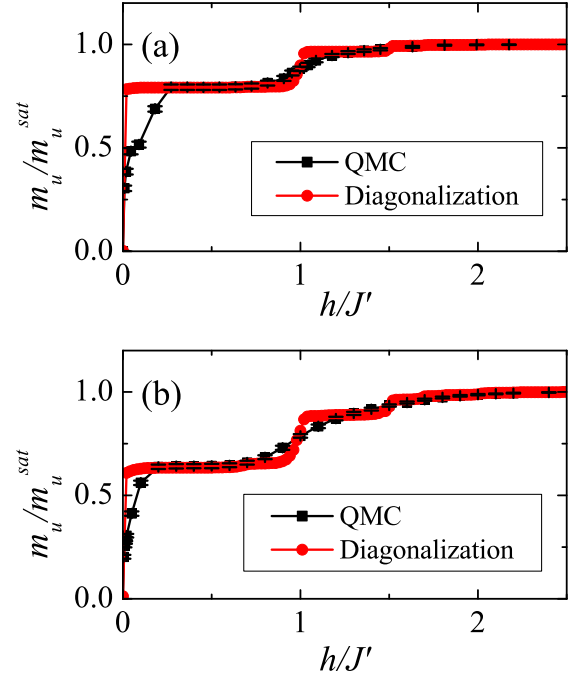}
\caption{Field dependence of the normalized magnetization
$m_u/m_u^{sat}$ in the effective model with $\xi=1.0$ and $p=1/16$
(in (a)) and $p=1/8$ (in (b)). The results obtained by
Eq.~\ref{e.mu_cluster} (red symbols) for a $L=100$ lattice are compared
with those from QMC simulations (black symbols) for a $L=40$
lattice.}\label{f.MuJP}
\end{center}
\end{figure}

We then compare the magnetization curve obtained by diagonalizing LM
clusters and then applying Eq.~\ref{e.mu_cluster} with the one via
QMC for the same effective model. The results are presented in
Fig.~\ref{f.MuJP} for two values of dilution concentration. We see that
the curves obtained by the two different methods agree quite well, especially for
the heights and positions of PPs. This agreement confirms that the
PPs are ascribed to the distribution of small LM clusters. We notice
that the curves from the two methods do not match well for very
small $h$, and at $h/J'\approx1$. These discrepancies are closely
related to the two assumptions for the LM clusters: since we assume
all the single-site LMs are fully polarized by the magnetic field,
the first PP calculated by the diagonalization of clusters is
expected to appear at very low field. Moreover the interactions
among LM clusters, which were neglected in the diagonalization
method, lead to smoother transitions between PPs, and to
a significant smearing of the second PP. From Fig.~\ref{f.MuJP}(b), we see that
these features are captured by the QMC simulations, but they are missing
in the diagonalization method.

\subsection{Bose glass nature of the DLM phase}
\label{ss.BG-DLM}

  As seen in the previous subsection, some of the most
relevant features of the magnetization curve in the DLM
phase can be captured by a model of independent clusters
which are storing some LMs not aligned with the field.
Upon a standard spin-boson transformation for $S=1/2$
spins, which maps \emph{down}-spins onto bosons 
(hereafter called \emph{LM quasiparticles}) 
and \emph{up}-spins onto bosonic holes
($S^+\rightarrow b$, $S^-\rightarrow b^{\dagger}$,
and $S^z\rightarrow 1/2-b^{\dagger}b$),
the LM Hamiltonian, Eq.~\eqref{e.ham_LRspin},
takes the following form
\begin{equation}
H=-\sum_{i<j} t_{ij} (b^\dag_i b_j + b^\dag_j b_i) + \sum_{i<j}
V_{ij}n_{i}n_{j} - \sum_i \mu_{i}n_{i}, \label{e.ham_bosons}
\end{equation}
with parameters
\begin{eqnarray}
t_{ij}=\frac{1}{2}|J^{\rm (eff)}_{ij}|~; \\ \label{e.t_bos} V_{ij}=J^{\rm (eff)}_{ij}~;\\
\label{e.V_bos} \mu_i= -h-\frac{1}{2}\sum_{j}V_{ij}~. \label{e.mu_bos}
\end{eqnarray}

The picture of the DLM phase as dominated by strongly
quantum fluctuating LM clusters can be then recast
in the bosonic language as a phase in which LM quasiparticles 
are expelled almost completely from the
isolated sites, and they remain \emph{localized}
on the clusters, namely they resonate between
two sites on a LM dimer, they localize around
the central site of a LM trimer, etc.
 We then apply the cluster decomposition of the
Hamiltonian introduced in the previous subsection,
and we neglect inter-cluster couplings in the
dilute limit. Neglecting the coupling between
clusters and isolated spins as well, we
obtain a ground state with isolated sites
completely empty of bosons, and clusters hosting
one or more boson. Transfering a bosons from a
cluster $C$ to an isolated site has a chemical potential
price $\Delta\mu \sim \Delta_C \sim J'$ where $\Delta_C$
is the gap to full polarization (or boson depletion)
of the cluster. Given that all couplings between
the cluster and the isolated spins are smaller than
$J'$ we can introduce them perturbatively. In doing so,
we can still neglect the density-density coupling
$V_{ij}n_{i}n_{j}$ with $i\in C$ and $j\in \bar C$,
because the bosonic occupation of the isolated
spins is negligible. The hopping terms $\sim t_{ij}$
for $i\in C$ and $j\in \bar C$ lead in turn to
a perturbative correction to the wave function of
LM quasiparticles, which acquires an \emph{exponentially}
decaying tail (with localization length
$\sim |\ln(J_{\rm ave}/J')|^{-1}$)
over the nearest isolated sites,
as it usually happens for a particle in a
potential well of depth $\sim \Delta_C$.

  The resulting picture of exponentially localized
LM quasiparticles around the LM cluster locations is hence
that of a \emph{Bose glass}. The
density of states of the exponentially localized
bosonic states is continuous, and correspondingly
an infinitesimal change of the applied field
always leads to adding/eliminating one boson,
which is a characteristic feature of a Bose glass.
This picture will become evident when discussing
the complete phase diagram of the system, as shown
in the following section.

\section{Phase Transitions in the 2D Site-Diluted Coupled Dimer System}
\label{s.PD_entire}

 In this section, we discuss the complete $T=0$ phase diagram of
dimer systems with general coupling ratios
$0\leqslant J'/J\leqslant1$. This will allow us to give a
unified picture of the novel quantum-disordered phases
introduced by site dilution in the weakly coupled dimer
systems. The main result is that the DLM phase, discussed
in details in the previous section, is continuously
connected to the Bose glass phase at higher fields.
This is further evidence of the Bose glass nature
of the DLM phase. \cite{Roscilde06,RoscildeH05, RoscildeH06}

\subsection{Phase Diagram with Site Dilution}

\begin{figure}[h]
\begin{center}
\null~~~~~~\includegraphics[
bbllx=23pt,bblly=31pt,bburx=563pt,bbury=455pt,%
height=70mm,angle=0]{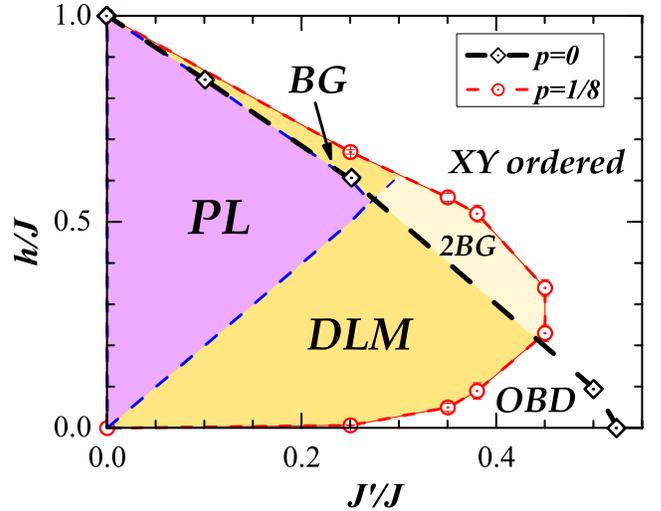}
\caption{Phase diagram of a site-diluted coupled dimer system at a dilution
concentration $p=0.125$. 
Three distinct phases are shown: the gapped plateau (PL) phase, 
corresponding to a bosonic Mott insulator; the disordered-local-moment (DLM) phase, corresponding to a Bose glass of LM quasiparticles;
Bose glass (BG) phase of dimer triplets, and the two-BG phase, corresponding
to a coexistence of LM-quasiparticle and dimer-triplet BG; and the XY ordered
phase, corresponding to a condensate of magnetic quasiparticles.   
The phase boundary in the clean limit $p=0$ (black dashed line) is
also shown for comparison.} \label{f.PhDp0.125}
\end{center}
\end{figure}

We start by briefly recalling the $T=0$ phase diagram of a
coupled-dimer system in the clean limit. At zero field, the ground
state is quantum disordered and gapped for weak inter-dimer couplings,
and it turns into a N\'eel-ordered state when the ratio $J'/J$ crosses the quantum
critical point at $(J'/J)_c \approx 0.523$.~\cite{Matsumotoetal01}
This critical point belongs to the universality class of the
classical three-dimensional Heisenberg model,~\cite{Matsumotoetal01,
Chen93} with $\nu=0.71(1)$, $\beta=0.36(1)$, and $z=1$. In the
presence of a magnetic field, the critical point at zero field
extends to a critical curve in the 2D $(h/J)$-$(J'/J)$ plane (see
dashed line in Fig.~\ref{f.PhDp0.125}). This curve separates the
quantum disordered phase at low field and low inter-dimer coupling
from the AFLRO phase. For this two-dimensional system, the universality
class of this transition is the mean-field one ($\nu=1/2, \beta=1/2$, etc.)
with $z=2$. \cite{Fisheretal89, Sachdev99}

Fig.~\ref{f.PhDp0.125} shows the complete phase diagram of the
two-dimensional doped  system with $p=1/8$.
As already mentioned in the introduction, doping with a small
concentration of non-magnetic impurities immediately induces
antiferromagnetic order in the zero-field
ground state for every finite inter-dimer coupling $J'$.
This is in agreement with the mean-field approach of Ref.~\onlinecite{Mikeskaetal04}.
For a large interval of $J'/J \lesssim 0.45$ (comprising the
case $J'=J/4$ of Sec.~\ref{ss.PD_weak}),
the system is driven back to a quantum disordered phase upon application
of a field $h=h_{\rm DLM}\ll J'$.
At a
higher field $h_{\rm BG}$ $(>\Delta_0)$, the system experiences
a second quantum
phase transition back to the ordered phase, as
discussed in Refs.~\onlinecite{RoscildeH05,RoscildeH06,Roscilde06}.
For $J'/J\lesssim 0.275$ 
the disordered phase is actually
a multiple one, composed of two gapless regions - a
DLM phase at low fields and a BG phase at higher field
\cite{RoscildeH05,RoscildeH06,Roscilde06} - and an intermediate
gapped plateau phase. The plateau phase is bounded from above
by the critical line of the clean system, as expected
from the fact that below that line no dimer triplet can
appear on the rare regions of intact dimers.
The curve that bounds the plateau phase from below, on
the contrary, corresponds to $h\simeq 2J'$;
this is also to be expected as the lower critical field for
this phase corresponds to the polarization field for
all LM clusters whose inter-LM coupling is $J'$,
and whose effective polarization field is effectively
given by $2J'$ (see also the discussion in 
Sec.~\ref{ss.Magnetization} and in Ref.~\onlinecite{noteonclusters}).
Hence the two curves bounding the plateau phase
from above and from below have opposite slopes, and
therefore cross each other at $J'/J\lesssim 0.275$, 
where the plateau phase disappears from the phase
diagram, leaving space to a unique, continuous
\emph{gapless} disordered phase
for $0.275 \lesssim J'/J\lesssim 0.45$.

 Finally, for $J'/J\gtrsim 0.45$ the low-field OBD
 phase merges with the high-field superfluid phase,
 marking the disappearance of all quantum-disordered
 phases in the system. It is remarkable to observe
 that this happens well below the critical value
 $J'/J = 0.523$ which marks the disappearance
 of the quantum-disordered phase in the clean
 limit. Hence site dilution of the lattice has the
 effect of significantly
 shifting the critical $J'/J $ ratio for the existence
 of a disordered phase in the system, as observed
 experimentally in Mg-doped TlCuCl$_3$. \cite{Fujisawaetal06}

 The topology of the phase diagram of Fig.~\ref{f.PhDp0.125} is generic
 for lattices of weakly coupled antiferromagnetic dimers (without frustration)
 with a concentration of vacancies that falls below the percolation threshold
 of the lattice. In particular this phase diagram will apply both to$2D$ and to $3D$
 arrays of dimers, which are percolating up to a finite concentration of vacancies.
 In particular a fundamental feature is that the disordered and gapless
 phases (DLM, 2BG and BG) are simply connected in the phase diagram,
 so that they systematically separate the gapped phase of fully polarized
 LMs from the magnetically ordered phases. In other words, in presence of
 site dilution there is \emph{no direct transition} between a spin-gapped phase
 and a magnetically ordered phase. This particular aspect of the topology
 of the phase diagram is a general property of that of bosons in a disordered
 lattice \cite{Fisheretal89, Gurarieetal09}, with the correspondence between
 spin-gap phases and Mott-/band-insulating phases, and between
 magnetically ordered phases and superfluid phases.

 When considering vacancy concentrations larger than that studied
 here (but still below the percolation threshold), one expects the gapless phases
 (DLM and Bose glass) to grow
 at the expenses of the plateau phase. The magnetically ordered phase
 at low fields (OBD phase) will also grow towards larger fields, as a denser network of LMs
 develops a magnetic order which is more robust to the application of a field.
 On the other hand, the ordered phase at large fields will instead shrink at the
 expenses of the Bose glass phase, given that a larger triplet density
 (namely a larger magnetization) is necessary for the triplet gas to condense
 when the disorder is higher.
 It is important to contrast the phase diagram of Fig.~\ref{f.PhDp0.125} with
 the one presented in Ref.~\onlinecite{Mikeskaetal04} on the basis of mean-field
 calculations. In the mean-field phase diagram all the gapless disordered phases,
 dominated by Anderson localization of triplets and/or moments aligned oppositely
 to the field, are completely absent. This is easily understandable since Anderson
 localization cannot be captured at the mean-field level; the gapless nature
 of the above mentioned phases comes from \emph{rare}, extended regions which
 can host gapless excitations, while the \emph{typical} behavior (captured at the
 mean-field level) is that of a gapped system.

\subsection{Two-species Bose glass}
\label{s.2BG}

 As seen in the previous subsection,
the DLM phase and the BG phase are indeed continuously
connected, and the DLM phase has a nature which is
analogous to that of the BG phase. In
Section \ref{ss.BG-DLM} we pointed out that the DLM
phase is characterized by exponentially localized
quasiparticles corresponding to LMs antiparallel
to the field, while the BG phase is characterized by
the exponential localization of dimer triplets.
For $0.275 \lesssim
J'/J\lesssim 0.45$ localized quasiparticles
of \emph{both} species appear above the transition curve
for the clean system (which is the condition for
the presence of dimer triplets). The
appearance of localized dimer triplets
\emph{before} the LM quasiparticles localized on
LM clusters disappear upon increasing the field
guarantees the gapless nature of the many-body spectrum
throughout the quantum disordered regime.
Hence the phase diagram region for $J'/J\gtrsim 0.275$
comprised between the two curves $h_{\rm BG}(J'/J)$
for the doped system and  $h_{\rm c}(J'/J)$
for the clean system represents a \emph{two-species} Bose glass
(2BG).

\begin{figure}[h]
\begin{center}
\includegraphics[
bbllx=40pt,bblly=40pt,bburx=700pt,bbury=600pt,%
     width=80mm,angle=0]{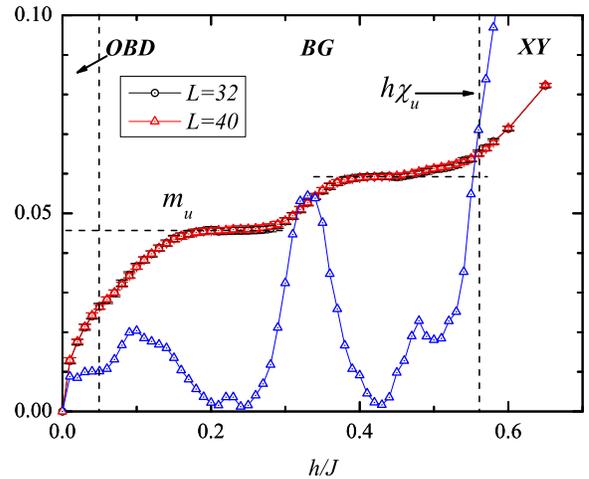}
\caption{Field evolution of the uniform magnetization $m_u$ and the
uniform susceptibility $\chi_u$ in the site-diluted dimer systems
with $J'/J=0.35$ and $p=1/8$. Although there is no true
magnetization plateau, two pseudo-plateaus $m_u/m_u^{\rm
sat}\approx0.73$ and $m_u/m_u^{\rm sat}\approx0.95$ are resolved.}
\label{f.muJ0.35}
\end{center}
\end{figure}

  Fig.~\ref{f.muJ0.35} shows the magnetization curve
along a section of the phase diagram ($J'/J=0.35$)
crossing the 2BG phase. Similarly to the case
of weaker inter-dimer coupling thoroughly
investigated in this paper, clear PPs, corresponding
to LM dimers and trimers, are observed at
$m_u/m_u^{\rm (sat)} \approx 0.73$ and $m_u/m_u^{\rm (sat)}\approx 0.95$,
where $m_u^{\rm (sat)}=p/2$ is the saturation magnetization of the
LMs. The plateau at $m_u^{\rm (sat)}$ is on the contrary completely
removed by the appearance of dimer triplets at a field
$h_c(J'/J)\approx 0.4 J$, which marks the onset of the
2BG phase.

\subsection{Quantum Scaling and Critical Exponents}\label{ss.Scaling}

 Next we investigate the properties of the quantum critical
line separating the quantum disordered phase from the ordered phase
in the phase diagram of Fig.~\ref{f.PhDp0.125}. As already pointed
out above, for weak inter-dimer coupling ($J'\ll J$) this critical line divides ordered and disordered 
phases which have a different nature when considering the low-field or the intermediate-field
region. At low fields, the critical field ($h_{\rm DLM}$) divides the OBD phase from a
DLM phase, while at higher fields, the critical field ($h_{\rm BG}$) divides a
BG phase of dimer triplets from a condensate of those triplets
(corresponding to an XY ordered antiferromagnetic
phase). Hence a fundamental question
arises whether the quantum critical curve is characterized by
universal critical exponents or not.

\begin{figure}[h]
\begin{center}
\null~~~~~~\includegraphics[
clip,width=80mm,angle=0]{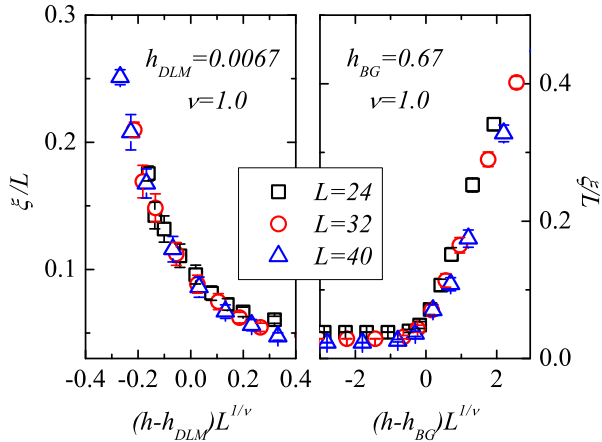}
\caption{Finite size scaling of the correlation length in the
vicinities of the critical fields $h_{\rm DLM}$ (left) and $h_{\rm BG}$
(right). The inter-dimer coupling is $J'=J/4$, and the dilution
concentration is $p=0.125$. } \label{f.Xiscale}
\end{center}
\end{figure}

\begin{figure}[h]
\begin{center}
\null~~~~~~\includegraphics[
clip,width=80mm,angle=0]{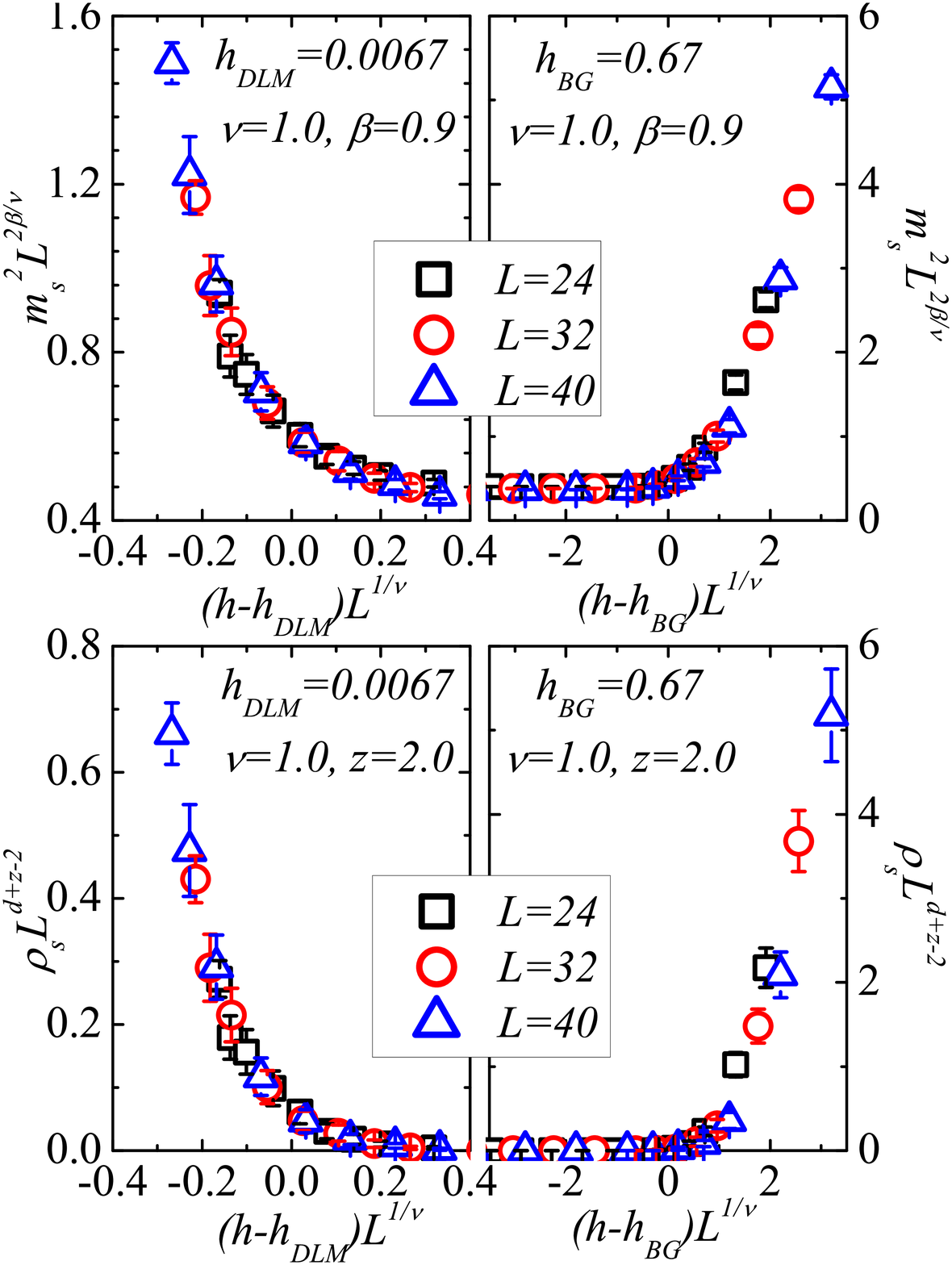}
\caption{Finite size scaling of the staggered magnetization and the
spin stiffness in the vicinities of the critical fields $h_{\rm DLM}$
(left) and $h_{\rm BG}$ (right). The model parameters are the same
as in the caption of Fig.~\ref{f.Xiscale}.} \label{f.Msscale}
\end{center}
\end{figure}

We answer this question by a detailed numerical
scaling analysis of the transition line.
 The correlation length is extracted from the transverse
structure factor Eq.~\eqref{e.ssf} by a second-moment
method,~\cite{Cooperetal82, Matsumotoetal01} assuming that
$S^\bot(\vec{k})$ follows a Lorentzian shape close to
$\vec{k}=(\pi,\pi)$:
\begin{equation}
\xi=\frac{1}{\Delta k}\sqrt{\frac{S^\bot(\pi,\pi)}{S^\bot(\pi+\Delta
k,\pi)}-1},
\end{equation}
where $\Delta k=\frac{2\pi}{L}$. The scaling form of the correlation
length is
\begin{equation}
\xi=Lf_\xi[L^{1/\nu}(h-h_c)],\label{e.Xiscale}
\end{equation}
where $f_\xi[\cdot]$ denotes the corresponding scaling function.
This defines the correlation exponent $\nu$. The scaling plots
of the correlation length for $J'/J=1/4$ around the two critical
fields $h_{\rm DLM}=0.007(1)$ and $h_{\rm BG}=0.67(1)$
are shown in Fig.~\ref{f.Xiscale}, and they both show
the best collapse for an exponent $\nu \approx 1.0(1)$.

 To further characterize the universality class of the
transition we consider the scaling properties of the
staggered magnetization and spin
stiffness, whose scaling forms are
\begin{eqnarray}
m_s=L^{-\beta/\nu}f_{m_s}[L^{1/\nu}(h-h_c)],\\\label{e.Msscale}
\rho_s=L^{-(D+z-2)}f_{\rho_s}[L^{1/\nu}(h-h_c)].\label{e.Rhosscale}
\end{eqnarray}
Here $\beta$ is the order parameter exponent, $z$ is the
dynamical exponent, and $D=2$ is the spatial dimension of the
system. Fig.~\ref{f.Msscale} shows the associated scaling plots,
which give best collapse for critical fields $h_{\rm DLM}$
and $h_{\rm BG}$ and for a correlation exponent $\nu$
all consistent with the values estimated via the scaling of
the correlation length. Moreover we extract the other exponents
$\beta=0.9(1)$ and $z=2.0(1)$ at both $h_{\rm DLM}$ and
$h_{\rm BG}$.

\begin{figure}[h]
\begin{center}
\includegraphics[
clip,width=80mm,angle=0]{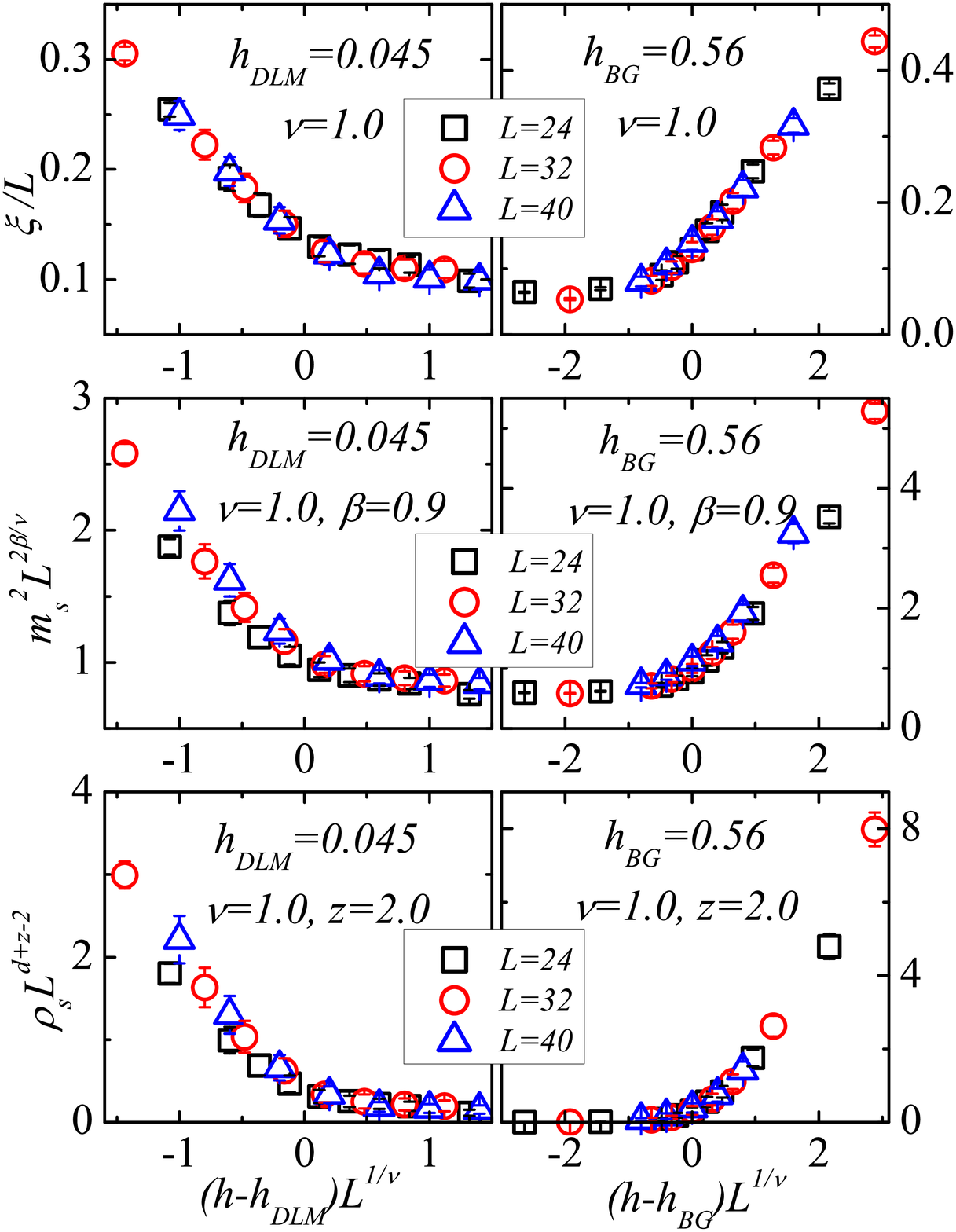}
\caption{From top to bottom: finite size scaling of correlation
length $\xi$, staggered magnetization $m_s$, and spin stiffness
$\rho_s$ at critical fields $h_{\rm DLM}=0.05(1)$ (left) and $h_{\rm
BG}=0.56(1)$ (right) in the site-diluted coupled dimers with
$J'/J=0.35$ and $p=1/8$.} \label{f.Scale0.35}
\end{center}
\end{figure}

For a BG-SF transitions, it was theoretically predicted that
$z=D=2$,~\cite{Fisheretal89} in agreement with our results. Moreover
for a well defined transition to occur in a disordered system the
correlation length exponent $\nu$ must satisfy the Harris
criterion~\cite{Chayesetal86} $\nu\geqslant2/D=1$, which is also
verified by our results. Repeating the same scaling analysis along
the critical line of the phase diagram we obtain analogous estimates
for the critical exponents, as shown explicitly in
Fig.~\ref{f.Scale0.35} for the case $J'/J=0.35$; this is indeed
a non-trivial result, as the BG phase close to $h_{\rm BG}$ in this
case is a 2BG, as discussed in Section~\ref{s.2BG}, and across the
transition at  $h_{\rm BG}$ the dimer triplets condense, while the LM
quasiparticles remain localized.

 Hence all the above results point towards a critical
curve with \emph{universal} critical exponents defining the SF-BG quantum
phase transition in $D=2$. The exponents $\beta$, $\nu$
and $z$ are fully consistent with those estimated in
Ref.~\onlinecite{Roscilde06} for the SF-BG transition in a
differently coupled-dimer system (a strongly coupled bilayer)
with site dilution, and therefore we argue that they
apply to the condensation transition of a generic
two-dimensional dirty-boson system with incommensurate
particle density.

\section{Conclusions}\label{s.Conclusion}

Making use of Stochastic Series Expansion Quantum Monte Carlo,
we systematically investigated the zero-temperature
phase diagram of weakly coupled dimers of $S=1/2$ spins with site dilution and
in the presence of a magnetic field. We focused our attention on the specific case
of a planar array of coupled dimers.
In particular we find that the phase diagram is much richer than the clean
system, and also richer than what has been predicted at the mean-field level
\cite{Mikeskaetal04}. In particular, we show that the impurity-induced
ordered state of localizes moments (LMs) at zero field is destroyed by a tiny field
for sufficiently small dilution, and that the field drives the system to a novel,
disordered-local-moment (DLM) phase, which extends over a very
large field range. We elucidate the microscopic nature of this
gapless and quantum-disordered phase by comparing results
for the diluted coupled-dimer array with those obtained for a
network of $S=1/2$ LMs interacting via
effective couplings exponentially decaying with the distance.
We observe that the DLM phase is characterized by the fact
that spins pointing oppositely to the field remain \emph{localized}
on rare clusters of nearest-neighboring LMs. A spin-to-boson mapping
shows that the DLM phase is akin to a Bose glass phase of
localized hardcore bosons. In the DLM phase, the magnetization curve
of the system shows a hierarchy of pseudo-plateaus marking each
the full polarization of a class of LM clusters (dimers, trimers, etc.).
The specific quantum nature of the DLM phase is revealed by
direct comparison with data for the classical limit of the LM network.

 A systematic study of the phase diagram of the site-diluted
 planar array of dimers, as a function of the ratio of inter-dimer coupling
 to intra-dimer coupling and of the strength of the applied
 field, reveals that the DLM phase is continuously connected
 to the Bose glass phase occurring at higher fields, and
 associated with the Anderson localization of triplet quasi-particles
 appearing on the intact dimers. This continuous disordered and gapless
 phase separates the gapped phase of fully polarized
 LMs from the ordered phases -- either the impurity-induced phase
 present also at zero field, or the field-induced ordered phases characterized
 by condensation of the triplet quasi-particles. This special feature
 of the phase diagram is consistent with what expected for
 dirty-boson systems on a lattice. Moreover the critical
 exponents for the quantum phase transition separating the gapless
 disordered phases and the ordered phases appear to be same regardless
 the specific nature of the ordered and disordered
 phases which are connected by it. This signals the existence of a
 novel universality class for the Bose glass-to-superfluid transition
 in 2D  with $\nu\approx 1.0$, $z\approx D=2$, and $\beta \approx 0.9$.

 The above findings show that the observation of unconventional
disordered phases without a gap, induced by disorder, is quite
realistic in the context of doped spin-gap antiferromagnets.
In fact, recent studies have focused on the properties of the
bond-disordered spin ladder IPA-Cu(Cl$_x$ Br$_{1-x}$)$_3$
\cite{Gotoetal08, Manakaetal08, Manakaetal09, Hongetal10},
giving the first evidence of a magnetic Bose glass in this
system \cite{Manakaetal09, Hongetal10}. In the context of
bond disorder the Bose glass appears as a consequence of
localization of triplets induced by intense fields \cite{Nohadanietal05},
while no DLM phase is expected. On the other hand, site dilution of the magnetic
lattice, realized by non-magnetic doping, would realize a much
broader Bose glass region (comprising the DLM phase), to be
observed also at small magnetic fields. This is not at all a negligible
aspect: indeed the smallness of the field for which the DLM phase
sets in (which can be two or more orders of magnitude smaller than the dominant
antiferromagnetic coupling) makes the DLM phase observable also in compounds
which have a large spin-gap (of the order of 10 K or higher),
and for which the physics of spin-triplet condensation is not
easily accessible in experiments. A possible candidate
compound is the recently investigated spin ladder compound
Bi(Cu$_{1-x}$Zn$_x$)$_2$PO$_6$, \cite{Bobroffetal09}
featuring a spin gap of $\sim 35$ K with zero doping ($x=0$).
Other potential candidates include
the spin-ladder compounds $\rm{(C_5H_{12}N)_2CuBr_4}$ \cite{Chaboussantetal97,
Watsonetal01, Klanjseketal08}
and  IPA-CuCl$_3$ \cite{Masudaetal06} with Zn of Mg doping
of the Cu ions.

\section{Acknowledgement} We thank N. Bray-Ali, L. Ding,
T. Giamarchi, W. Li, G. Misguich, P. Sengupta, M. Sigrist, S. Wessel
for fruitful discussions. This work is supported by the
Department of Energy through grant No. DE-FG02-05ER46240.
Computational facilities have been generously provided by the
high-performance computing center at USC.



\end{document}